\documentclass[11pt]{article}
\usepackage{graphicx}
\usepackage{epsfig}
\usepackage{epsf,amsfonts,amssymb}
\newcommand{\be}{\begin{equation}}
\newcommand{\ee}{\end{equation}}
\newcommand{\bea}{\begin{eqnarray}}
\newcommand{\eea}{\end{eqnarray}}
\newcommand{\ba}{\begin{eqnarray}}
\newcommand{\ea}{\end{eqnarray}}

\newcommand{\beq}{\begin{equation}}
\newcommand{\eeq}{\end{equation}}
\newcommand{\beqa}{\begin{eqnarray}}
\newcommand{\eeqa}{\end{eqnarray}}
\newcommand{\beqar}{\begin{eqnarray*}}
\newcommand{\eeqar}{\end{eqnarray*}}

\newcommand{\reef}[1]{(\ref{#1})}

\newcommand{\eg}{{\it e.g.,}\ }
\newcommand{\ie}{{\it i.e.,}\ }

\newcommand{\sech}{\mathrm{sech}\ }
\def\O{\mathcal{O}}

\begin{document}

\setlength{\unitlength}{1mm}

\thispagestyle{empty}
 \vspace*{2.5cm}

\begin{center}
{\bf \LARGE Microscopic Theory of Black Hole Superradiance}\\

\vspace*{2cm}

{\bf \'Oscar J.~C.~Dias,}$^1\,$ {\bf Roberto Emparan,}$^{1,2}\,$
{\bf Alessandro Maccarrone,}$^1\,$

\vspace*{0.5cm}

{\it $^1\,$Departament de F{\'\i}sica Fonamental, Universitat de
Barcelona, \\
Marti i Franqu{\`e}s 1,
E-08028 Barcelona}\\[.3em]
{\it $^2\,$Instituci\'o Catalana de Recerca i Estudis Avan\c{c}ats (ICREA)}\\[.3em]

\vspace*{0.3cm} {\tt odias@ub.edu, emparan@ub.edu, sandro@ffn.ub.es}

\vspace*{2cm}

\vspace{.8cm} {\bf ABSTRACT}
\end{center}

We study how black hole superradiance appears in string microscopic
models of rotating black holes. In order to disentangle
superradiance from finite-temperature effects, we consider an
extremal, rotating D1-D5-P black hole that has an ergosphere and is
not supersymmetric. We explain how the microscopic dual accounts for
the superradiant ergosphere of this black hole. The bound
$0<\omega<m\Omega_H$ on superradiant mode frequencies is argued to
be a consequence of Fermi-Dirac statistics for the spin-carrying
degrees of freedom in the dual CFT. We also compute the superradiant
emission rates from both sides of the correspondence, and show their
agreement.

\noindent

\vfill \setcounter{page}{0} \setcounter{footnote}{0}
\newpage

\tableofcontents

\setcounter{equation}{0}

\setcounter{equation}{0}\section{\label{sec:Introduction}Introduction}

The microscopic string theory of black holes provides an accurate
statistical counting of the Bekenstein-Hawking entropy
\cite{Strominger:1996sh}-\cite{Horowitz:2007xq}
and a microscopic picture of Hawking radiation
\cite{Das:1996wn}-\cite{Gubser:1997qr}
at least for some classes of black holes. In the present paper we
address how this microscopic theory also accounts for a
characteristic phenomenon of rotating black holes: the black hole
superradiance.

Superradiance is a phenomenon associated to the presence of an
ergoregion around the black hole \cite{zel}-\cite{Cardoso:2004nk}
Since the Killing
vector that defines the energy measured by asymptotic observers
becomes spacelike within the ergosurface, it follows that in the
ergoregion there can exist excitations with negative energy relative
to infinity. So if we scatter a wave off the black hole, this wave
can excite negative energy modes that may subsequently fall into the
horizon. To an asymptotic observer this will appear as a positive
energy flux coming out of the horizon, and the scattered wave can
emerge with higher amplitude than the impinging wave: this is known
as superradiant scattering. If an incident wave $\Phi\sim
f(r,\theta) e^{-i\omega t+i m \phi}$, with energy $\omega>0$ and
angular momentum number $m$, scatters off a black hole with horizon
angular velocity $\Omega_H$, the requirement that a negative-energy
flux crosses the horizon towards the future is
 \beq\label{sradbound}
0<\omega< m\Omega_H\,.
 \eeq
Only modes satisfying this condition can undergo superradiant
amplification.

Superradiant scattering can be regarded as stimulated emission, and,
just like the latter (classical) process is related by detailed
balance to (quantum) spontaneous emission, rotating black holes are
also known to spontaneously emit superradiant modes within the range
\reef{sradbound}, in a process closely related to Hawking radiation.
These carry away energy, but also angular momentum off the black
hole. In our microscopic picture it is convenient to first describe
the process of spontaneous superradiant emission, and then infer the
stimulated emission.

When the black hole temperature is different from zero it is
difficult to disentangle spontaneous superradiant emission from
thermal Hawking radiation---in fact both become part of one and the
same phenomenon. In this paper, however, we are interested only in
the microphysics behind the presence of an ergoregion and the
existence of superradiant modes \reef{sradbound}. So we will
investigate the spontaneous emission from an extremal, \ie
zero-temperature, rotating black hole, for which thermal Hawking
radiation is absent. Since the black hole has a `cold' ergoregion,
we refer to it as an {\it ergo-cold black hole}. This will allow us
to isolate superradiance: only modes that satisfy \reef{sradbound}
will be emitted. Note, however, that after the emission of
superradiant quanta begins, the angular momentum will be reduced
below its maximal value and the black hole will be driven away from
extremality, so thermal Hawking radiation will promptly set in. It
is the onset of the decay that will give us more neatly the
microscopic basis of the superradiant bound \reef{sradbound}.

There have been previous papers dealing with emission rates from
rotating black holes and the microscopic calculations that match
them \cite{Maldacena:1997ih,Cvetic:1997uw,Cvetic:1998xh,
Bardeen:1999px} (see \cite{Gubser:1998ex,David:2002wn} for a
review), in some cases discussing, more or less directly, aspects of
superradiance. Typically, these papers have computed the absorption
cross sections for a non-extremal black hole and for its microscopic
finite-temperature dual. Even if these results exhibit essential
agreement between both sides, we feel that the long calculations
involved, and the mixing with thermal Hawking radiation, hide some
very simple microphysics behind \reef{sradbound}. We hope to clarify
the microscopic origin of the ergoregion and provide a simple
interpretation of the superradiant modes in it. We shall follow
mostly a suggestion advanced in \cite{Emparan:2007en}, making it
more precise and quantitative. A salient conclusion of our analysis
is a clear understanding of the bound \reef{sradbound} as
essentially a consequence of Fermi-Dirac statistics for the
microscopic degrees of freedom that give the black hole its angular
momentum.

The paper is structured as follows. The main ideas are introduced
first in a fairly self-contained and elementary discussion, while
the technically most involved analysis is postponed to later
sections. So, section \ref{sec:bound} begins with a qualitative
review of the microscopic model of D1-D5-P black holes, with and
without ergospheres, and then proceeds to derive \reef{sradbound}
from simple microscopic considerations. The detailed calculations of
absorption rates, which are needed for other quantitative aspects of
superradiance, are studied at the supergravity level in
sec.~\ref{sec:Sugra}. This is an extension of previous analyses of
radiation from the D1-D5-P black holes studied at length in
\cite{Cvetic:1997uw,Cvetic:1998xh}. We do generalize their results
to include momentum for the bulk scalar. The microscopic side is
then developed in sec.~\ref{sec:dualcft}. Here we first establish
the details of the identification of the dual CFT state, compute the
microscopic absorption cross section, and compare to the
supergravity results. Sec.~\ref{sec:Conclusions} concludes with a
qualitative discussion of how our picture accounts for superradiance
in other systems with `cold ergoregions'. The appendix contains an
analysis of how the near-horizon geometry encodes information about
the possibility of superradiance in the full geometry.

\setcounter{equation}{0}\section{Microphysics of cold ergoregions}
\label{sec:bound}

We begin by introducing the microscopic picture of superradiance and
then provide a simple derivation and interpretation of the bound
\reef{sradbound} for the ergo-cold black hole.

\subsection{Qualitative microscopic origin of the ergoregion}

Our basic picture applies to any black hole that admits an
`effective string' description, \ie to which AdS$_3$/CFT$_2$ duality
applies\footnote{And even to some that may not, like in
\cite{Emparan:2007en}, although in this case the bound
\reef{sradbound} is recovered only up to numerical factors.}, but
for definitiness we focus, for the most part, on the D1-D5-P system,
which describes a class of near-supersymmetric five-dimensional
black holes. We shall begin by reviewing in qualitative terms the
microscopic picture of several kinds of D1-D5-P black holes.

The D1 and D5-branes form a bound state whose low-energy dynamics is
described by a 1+1-dimensional field theory along their common
worldvolume directions (the other four directions wrap a small $T^4$
or $K3$). It is a non-chiral conformal field theory (CFT) with
$(4,4)$ supersymmetry, \ie both the left- and right-moving sectors
are supersymmetric. Supersymmetry itself will not play any essential
role in our discussion, but the existence of fermionic excitations
in at least one of the two chiral sectors is important. For large
numbers $N_1$, $N_5$, of D1 and D5 branes, the central charge of
both sectors is $c=6N_1 N_5$. The CFT can have left- and
right-moving excitations, with levels $L_0$ and $\bar L_0$,
corresponding to open string excitations propagating along the
worldvolume of the branes. These give rise to a linear momentum $P$.

When the spatial direction along this D1-D5-P system is compactified
on a circle of size $2\pi R$ (much larger than the other compact
directions), we obtain a five-dimensional configuration. Typically,
the state corresponding to a black hole has both sectors populated
by thermal ensembles of excitations with temperatures $T_L$ and
$T_R$. If the two sectors interact only very weakly, the total
entropy, energy and momentum are $S=S_L+S_R$, $E=P_L+P_R$ and
$P=P_L-P_R$, with quantized momenta $P_{L,R}=N_{L,R}/R$. Since
$T_{L,R}^{- 1}=\left(\partial S_{L,R}/\partial
P_{L,R}\right)=2\left(\partial S_{L,R}/\partial E\right)_P$, it
follows that the actual temperature $T_H^{-1}=\left(\partial
S/\partial E\right)_P$ of the entire configuration is \beq\label{TH}
T_H^{-1}=\frac{1}{2}\left(T_L^{-1}+T_R^{-1}\right)\,. \eeq If any of
the two sectors is in a ground state (either $T_L$ or $T_R$ vanish),
the temperature of the entire system vanishes.

\begin{figure}[th]
\centerline{\includegraphics[width=12cm]{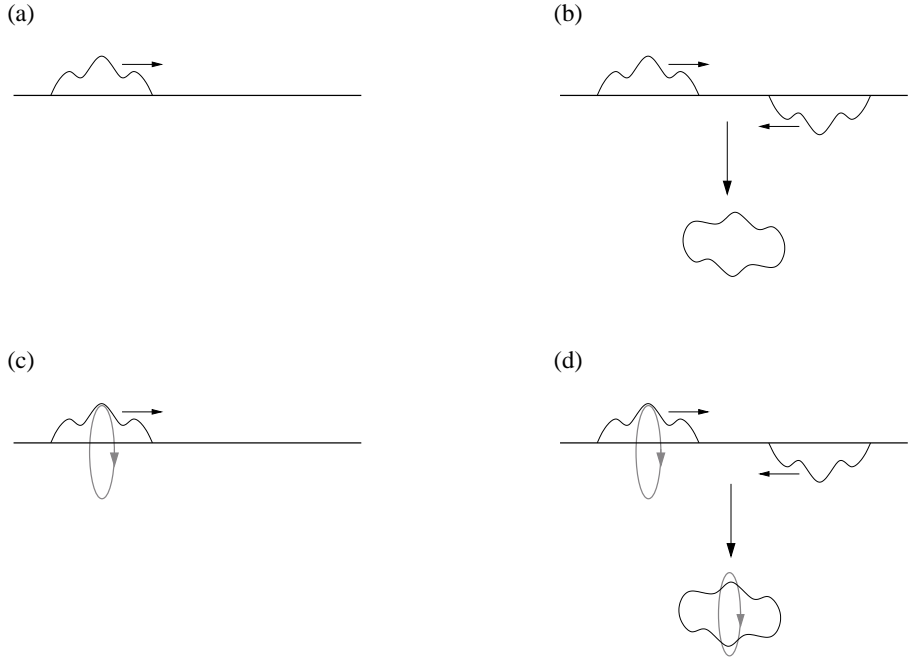}}
\caption{\small Four different kinds of black hole in the `effective
string' picture. The excitations of the two chiral sectors, with
levels $L_0$ (left-moving) and $\bar L_0$ (right-moving), correspond
to open strings attached to the brane bound state. (a) {\em
Supersymmetric static black hole}: $L_0= 0$, $\bar L_0=N_R$: only
the right-moving sector is excited. (b) {\em Near-supersymmetric
static black hole}: $ L_0=N_L> 0$, $\bar L_0=N_R> 0$. Left and
right-moving excitations can annihilate to emit a closed string:
this is Hawking radiation. (c) {\em Supersymmetric rotating black
hole}: $L_0=0$, $\bar L_0=N_R-6 J_R^2/c> 0$. The coherent
polarization of right-moving fermions yields a macroscopic
(self-dual) angular momentum $J_R$. In the absence of left-moving
open strings, there cannot be any radiation of closed strings, hence
there is no Hawking nor superradiant emission. (d) {\em Ergo-cold
black hole}: $ L_0> 0$, and $\bar L_0=N_R-6 J_R^2/c= 0$ with $N_R
>0$. The right-moving sector is a Fermi sea of polarized fermionic
excitations, so the temperature vanishes. Open strings in this
sector can interact with those in the left sector and emit closed
strings that carry angular momentum: {\it the black hole possesses a
superradiant ergosphere}. The superradiant bound on modes
\reef{sradbound} is directly related to the energy of the Fermi
level, and thus is a consequence of Fermi-Dirac statistics for the
excitations of the CFT.} \label{fig:sradiance}
\end{figure}

The simplest black hole corresponds to a thermal ensemble of
excitations in only one of the two sectors, say the right-moving
one. Supersymmetry of the left sector is then preserved, and
$T_L=T_H=0$. This is the static supersymmetric extremal black hole
of ref.~\cite{Strominger:1996sh}. If both sectors are excited, then
generically the system has $T_H\neq 0$. An open string excitation
from the left sector can combine with an open string from the right
sector, and form a closed string that propagates away into the bulk
of spacetime. This is the microscopic counterpart of Hawking
emission at temperature $T_H$ \cite{Callan:1996dv,Horowitz:1996fn}.

To include rotation, we take into account that the fermionic
excitations of the left and right sectors are charged under the
R-symmetry group $SU(2)_L \times SU(2)_R$ of the supersymmetric CFT.
These R-symmetries generate the five-dimensional spatial rotation
group $SO(4)\simeq SU(2)_L\times SU(2)_R$. So the R-charge
corresponds to spacetime angular momentum, $J_L$ or $J_R$,
respectively for left and right fermions. If many of these fermions
are coherently polarized we obtain a macroscopically large angular
momentum. This projection into definite polarization shifts the
levels as
\beq L_0=N_L-\frac{6J_L^2}{c}\,\qquad \bar
L_0=N_R-\frac{6J_R^2}{c}\,,
 \eeq
 and in particular the total entropy
and temperature are reduced.

Observe now that there are two distinct ways of achieving an
extremal ($T_H=0$) rotating black hole. In the first one we set,
say, $N_L=0=J_L$ (so half of the supersymmetry is preserved), $\bar
L_0>0$, and some of the right-moving fermions polarized to give
$J_R\neq 0$ \cite{Breckenridge:1996is}. Since only one of the two
sectors is excited, the left and right-moving open strings cannot
combine to emit a closed string. This fits nicely with the property
that the horizon of the corresponding black hole remains static
relative to asymptotic observers: since $\Omega_L=\Omega_R=0$ there
is no ergosphere nor superradiant emission, even if $J_R\neq 0$.

The second, less studied way to achieve a zero-temperature rotating
black hole is by having the right-moving sector contain only
polarized fermions that fill energy levels up until the Fermi level.
This occurs when
 \beq \label{coldergo}
 N_R=\frac{6J_R^2}{c}\,.
  \eeq
This is a ground state, $\bar L_0=0$, at fixed $J_R$, with zero
entropy and at zero-temperature. The left-moving sector is assumed
to be thermally excited, with $ L_0 >0$: this provides for the
entropy. Both sectors can carry angular momentum, so, in contrast to
the supersymmetric case, the total angular momentum need not be
self-dual nor anti-self-dual. More importantly, even if the system
is at zero temperature, both left and right moving open strings are
present and can annihilate to emit a closed string. Since the
right-moving open string necessarily carries spin, so will also the
emitted radiation. This is, qualitatively, what we expect from
superradiant emission. In fact, the corresponding black hole
possesses an ergosphere and superradiant emission is present. So we
have found a qualitative microscopic picture for the superradiance
from the ergo-cold black hole \cite{Emparan:2007en}.

\subsection{Microscopic derivation of the superradiant frequency bound}

We can be more quantitative and recover the superradiant frequency
bound from this microscopic picture. In five spacetime dimensions
the black hole can rotate in two independent planes and if we label
the rotation angles on these planes by $\phi$ and $\psi$ then the
bound \reef{sradbound} is generalized to \beq \label{sradbound2}
0<\omega< m_\phi\Omega_\phi+m_\psi\Omega_\psi\,, \eeq where
$\Omega_{\phi,\psi}$ are the horizon angular velocities on each
rotation plane, and $m_{\phi,\psi}$ the corresponding angular
momentum (``magnetic") quantum numbers. We may instead use the left
and right Euler angles $\psi_{L,R}=\phi\mp \psi$, in terms of which
the bound is \beq \label{sradbound3} 0<\omega<
m_L\Omega_L+m_R\Omega_R\,, \eeq with
$m_{L,R}=\frac{1}{2}\left(m_\phi\mp m_\psi\right)$ and
$\Omega_{L,R}=\Omega_\phi\mp \Omega_\psi$. This is slightly more
convenient, since as we saw above these angles diagonalize the
R-charges (\ie target-space spins) of the left and right-moving
fermions of the CFT.

The ergo-cold black hole described above has $\Omega_R\neq0$ and
$\Omega_L=0$ (although $J_L$ need not vanish). So the bound is \beq
\label{sradbound4} 0<\omega< m_R\Omega_R\,, \eeq \ie $m_L$ does not
limit the frequencies. We wish to derive eq.~\reef{sradbound4} from
our microscopic picture.

To begin with, we can easily obtain that at zero-temperature only
one of $\Omega_L$, $\Omega_R$, can be different from zero. The two
sectors of the CFT have negligible interaction, so
$S(E,P,J_L,J_R)=S_L(E_L,J_L)+S_R(E_R,J_R)$. For each sector we have
a chemical potential $\mu_{L,R}$ associated to the respective
R-charges, \ie $J_{L,R}$, through
 \beq
 \frac{\mu_{L,R}}{T_{L,R}}=-\left(\frac{\partial S_{L,R}}{\partial
J_{L,R}}\right)_{E_{L,R}}\,.
 \eeq
 The angular velocities of the total system are in
turn
 \beq \frac{\Omega_{L,R}}{T_H}=-\left(\frac{\partial
S(E,P,J_L,J_R)}{\partial J_{L,R}}\right)_{E,P}\,,
 \eeq
 where $T_H$ is the
total system's temperature \reef{TH}. Hence \beq\label{Omu}
\Omega_{L,R}=\frac{T_H}{T_{L,R}}\mu_{L,R}\,, \eeq and in the
extremal limit in which $T_R\to 0$, \beq \label{Omuext} \Omega_R\to
2\mu_R\,,\qquad \Omega_L\to 0\,. \eeq

As we explained above, for the ergo-cold black hole we take the
right sector of the CFT to be populated by polarized fermions
filling up to the Fermi level, so their number density distribution
is a step function \beq\label{fermirho}
\rho(\epsilon,j_R)=\Theta(j_R\mu_R-\epsilon)\,. \eeq Here $\epsilon$
is the energy and $j_R$ the R-charge of the fermion, \ie spin in
$SU(2)_R$, which in general can be $\pm 1/2$. We assume that in the
state \reef{fermirho} they are all polarized with $j_R=+1/2$, to
achieve maximum angular momentum, see (\ref{coldergo}). Using the
chemical potential $\mu_R$ introduced above, the Fermi energy is
\beq\label{fermien} \epsilon_{\rm
Fermi}=\frac{\mu_R}{2}=\frac{\Omega_R}{4}\,. \eeq

In this state it is possible to have a collision of left and
right-moving open strings creating a closed string massless scalar
mode. Our aim is to show that if this scalar has frequency $\omega$
and angular momentum numbers $\ell$, $m_R$ and $m_L$, then $\omega$
must lie in the range \reef{sradbound4}. In order for the scalar to
escape to infinity its energy must be positive, so we need only
derive the upper bound in \reef{sradbound4}.

The interaction vertex involves bosonic and fermionic open strings
from each sector, in either the initial or final states. But the
spin of the scalar is provided only by fermions. For a given $\ell$
the angular momentum of the scalar is in the $(\ell/2,\ell/2)$
representation of $SU(2)_L\times SU(2)_R$, \ie $|m_L|,|m_R|\leq
\ell/2$, so we need $\ell$ fermionic open strings from each sector
to match the spin quantum numbers of the scalar. A minimal scalar at
s-wave ($\ell=0$) couples to an operator of conformal dimension
$(1,1)$, typically of the form $\partial_+ X \partial_- X$, \ie one
boson from each sector. Then, at the $\ell^{\rm th}$ partial wave it
will couple to this boson pair and to the $\ell$ fermion pairs.
Additional bosons may be involved, but then the amplitudes are
suppressed by higher powers of the coupling and the frequency,
although we need not assume their absence.

For our system, the right-sector open strings in the initial-state
in the interaction can only be fermionic with $j_R=+1/2$. The
fermions in the final state can have either $j_R=\pm 1/2$: we take
the numbers of each kind of these to be $n_\pm$, so the number of
initial fermions from the right sector is $\ell-n_+-n_-$. The
balance of angular momentum in the interaction is then
 \beq
\frac{1}{2}\left(\ell-n_+-n_-\right)= m_R +\frac{1}{2}n_+ -
\frac{1}{2}n_-\,,
 \eeq
 \ie the closed string is emitted with
\beq\label{mr}
 m_R=\frac{\ell}{2}-n_+\,.
  \eeq
  We will not need to
consider any specific properties of the left-moving modes in our
analysis.

Both the left and right sectors contribute an equal amount
$\omega/2$ to the energy of the emitted closed string --- otherwise
the latter would carry the difference as a net momentum: this more
general case will be dealt with later below. The energy-budget of
the interaction in the right sector is then
\beq\label{budget}
\omega^{(f)\mathrm{in}}_R=\frac{\omega}{2}+
\omega^{(f)\mathrm{out}}_R+\omega^{(b)}_R\,,
 \eeq
where $f$ and $b$ denote fermionic and bosonic open strings. In the
lhs of this equation we have the energy of the $\ell-n_+-n_-$
initial fermions. Since their energy levels are bounded above by the
Fermi energy \reef{fermien}, we have
 \beq\label{omegafin}
\omega^{(f)\mathrm{in}}_R\leq (\ell-n_+-n_-)\epsilon_{\rm
Fermi}=(\ell-n_+-n_-)\frac{\Omega_R}{4}\,.
 \eeq
  As for the final
fermions, the energies of the $n_-$ fermions with $j_R=-1/2$ are not
constrained other than to be positive: they may fill states with
less or more energy than $\epsilon_{\rm Fermi}$. But the $n_+$
fermions with $j_R=+1/2$ must have energies above the Fermi level,
since in the initial configuration the levels below $\epsilon_{\rm
Fermi}$ are all filled with positive-spin fermions. This sets a
lower bound \beq\label{omegafout} \omega^{(f)\mathrm{out}}_R >
n_+\frac{\Omega_R}{4}\,. \eeq The energy of the bosonic open strings
is only constrained to be positive, $\omega^{(b)}_R>0$. Then,
eq.~\reef{budget}, together with \reef{mr}, \reef{omegafin} and
\reef{omegafout}, yields the inequality \beq\label{radderived}
\omega < m_R \Omega_R - \frac{n_-}{2}\Omega_R\leq m_R \Omega_R\,,
\eeq which reproduces exactly the superradiant bound
\reef{sradbound4} derived for the rotating black hole\footnote{The
bound is as close as possible to saturation when $n_-=0$, the boson
energy $\omega^{(b)}_R$ is minimal (set by the gap $\sim 1/N_1 N_5
R$), and all fermions are the closest possible to the Fermi energy
(\ie within $\sim 1/N_1 N_5 R$ of it). If $n_->0$ then this closest
value to the bound cannot be achieved.}. Note that this result
follows essentially from Pauli's exclusion principle for the
polarized fermions in the initial state: {\it the superradiant bound
on frequencies is a consequence of Fermi-Dirac statistics for the
carriers of angular momentum in the dual CFT.}

Note that at least one bosonic open string must appear in the
right-sector in the final state, so the system will not remain
extremal after it begins to radiate. This is also just like we
anticipated from the supergravity side.

The left-moving fermions, which can contribute arbitrarily to $m_L$,
have not played any role in this derivation. This is in accord with
the fact that when $\Omega_L=0$ (even if $J_L\neq 0$), $m_L$ does
not appear in the macroscopically-derived bound \reef{sradbound2}.

\subsection{Four-dimensional black holes}
\label{subsec:4dbhs}

This analysis applies almost immediately to the four-dimensional
black holes described by a dual chiral $(0,4)$ CFT. Only the right
sector is supersymmetric so the R-symmetry consists of a single
$SU(2)$ group. This corresponds to the four-dimensional rotation
group $SU(2)\simeq SO(3)$. Again, non-BPS extremal rotating black
holes exist, with four charges, that possess an ergosphere and the
accompanying superradiant modes satisfying \reef{sradbound}. The
dual microscopic state is essentially the same as above: the right
sector is filled up to $\epsilon_{\rm Fermi}=\Omega_H/4$ with
 fermions with $j=+1/2$, while the left sector is in a thermal ensemble and
accounts for the entropy. The emission of a closed string massless
scalar with quantum numbers $(\omega,\ell,m)$ involves $2\ell$
right-sector fermions since now $|m|\leq \ell$. So \reef{mr} is
replaced by \beq m=\ell-n_+\,. \eeq There is also one boson from the
right sector in the final state of the interaction. From the left
sector the only requirement is an operator of conformal dimension
$\Delta_L= 1+\ell$. Following the steps above we find
\beq\label{rad1derived} \omega < m \Omega_H -
\frac{n_-}{2}\Omega_H\leq m \Omega_H\,.
 \eeq
  Thus
eq.~\reef{sradbound} has been derived microscopically for this
ergo-cold black hole.

\subsection{No superradiant emission of linear momentum}

We can also consider the emission of closed strings that carry away
some of the momentum $P$ of the D1-D5-P system. This is also of
interest, as the momentum corresponds to one of the three charges of
the black hole and there is a charge-ergoregion associated to it.
From the six-dimensional perspective, the horizon of the black
string is moving with velocity $V_H$ along the string direction $y$,
and the superradiance bound for a mode $\sim \exp({-i\omega t +i p y
+i m_L\psi_L+im_R \psi_R})$ is modified to
 \beq \label{wp}
 p < \omega
<m_L\Omega_L+m_R\Omega_R+p V_H \,. \eeq
 In the non-BPS extremal
rotating limit that we study, the black hole has $\Omega_L\to 0$.
For a generic D1-D5-P black hole the velocity is $|V_H|\leq 1$, but
we are particularly interested in the decoupling limit in which the
D1 and D5 charges of the black hole are much larger than its
momentum or the energy above the BPS bound. In this limit, the
ergo-cold black hole has $V_H\to 1$, so the bound becomes
\beq\label{mombound} 0< \omega - p < m_R\Omega_R\,. \eeq

We can easily derive this again from microscopic considerations.
First note that the first law of thermodynamics gives \beq
\frac{V_H}{T_H}=-\left(\frac{\partial S}{\partial P}\right)_E\,.
\eeq Reasoning as we did when deriving \reef{TH} for a two-sector
system, we find
 \beq\label{VH}
V_H=\frac{T_H}{2}\left(T_R^{-1}-T_L^{-1}\right)=\frac{T_L-T_R}{T_L+T_R}\,,
\eeq
 so $V_H\to 1$ when $T_R\to 0$. Also observe that in any case
$|V_H|\leq 1$.

The left and right-moving open strings that interact to emit a
closed string of frequency $\omega$ and momentum $p$ do not in this
case have the same energy, but instead \beq
\epsilon_{L,R}=\frac{\omega \pm p}{2}\,. \eeq We can follow now the
same arguments for the right-sector dynamics that we used above,
only changing $\omega/2 \to \epsilon_R$. Hence we obtain \beq
\omega-p <m_R\Omega_R\,. \eeq In order to complete the derivation of
\reef{mombound} we need only notice that if the closed string is to
arrive at infinity as an on-shell, propagating state, it must
satisfy $\omega>0$ and $\omega^2-p^2\geq 0$, \ie $\omega\geq |p|\geq
p$.

This implies that there cannot be any superradiant emission of
linear momentum (\ie $P$ charge in five dimensions) unless angular
momentum is radiated as well. This is in spite of the fact that in
the black hole geometry there is a momentum ergoregion, even in the
absence of rotation. From the supergravity point of view, the reason
for this difference between the emission of linear and angular
momentum is that in the former case the contribution to the
effective potential for scalar propagation coming from the momentum
does not fall off at infinity but creates an asymptotic potential
barrier of height $p$, so if $\omega < |p|$ the wave is
asymptotically exponentially suppressed.

Put another way, in a KK reduction to five dimensions the scalar has
mass $|p|$ and a propagating wave at infinity must satisfy $\omega
>|p|$. So a would-be superradiant momentum mode, satisfying $\omega <
pV_H$, cannot escape to infinity since $V_H\leq 1$: if emitted, it
gets reflected back off to the black hole by the effective
potential. In contrast, the centrifugal potential barriers fall off
faster at large distances: the spin does not affect the dispersion
relation of the wave at infinity. From the microscopic perspective,
there is a possible interaction vertex for the emission of a scalar
with linear momentum and zero angular momentum: take an initial
state with only a left-moving boson, and a final state with a
right-moving boson and a bulk scalar. However, in this case the
scalar would have $\omega <|p|$ and therefore could only exist as a
virtual excitation\footnote{An alternative interpretation is in
terms of charge superradiance: an extremal Reissner-Nordstrom black
hole can spontaneously emit particles of charge $e$ and mass $m$
only if $|e|>m$ \cite{Gibbons:1975kk}. In our case, the 5D mass and
KK electric charge of the particles are both equal to $p$.}.

\subsection{Superradiant amplification, extremal and non-extremal}

We have obtained a microscopic picture for the spontaneous emission
of superradiant scalars off an extremal non-BPS rotating black hole
--- the ergo-cold black hole. It is clear now that, if there is an
incident flux of this scalar field on the black hole, then those
modes that satisfy the bound \reef{sradbound} will undergo
stimulated emission. This is simply the familiar phenomenon that the
amplitude to emit a boson is amplified by a factor $\sqrt{N+1}$ if
the state to which the system decays already contains $N$ bosons.
This is, superradiant amplification follows conventionally from the
relation between Einstein's $A$ and $B$ coefficients. For a
classical incident wave, \ie with large bosonic occupation number
$N$, the stimulated emission is then a classical process.

In more detail, in our system at zero temperature we have argued
that superradiant modes, and only them, can be emitted and have a
finite decay rate $\Gamma_{\ell m}(\omega)$. Moreover, the system
cannot absorb any superradiant mode: if in the argument that lead to
the superradiant bound \reef{sradbound4} we change the scalar from
the final to the initial state, \ie $\omega\to-\omega$, $m_R\to
-m_R$, we see that absorption of this scalar can only happen when
$\omega > m_R \Omega_R$. So, for an incident flux ${\mathcal F}_{\rm
in}$, detailed balance yields a total absorption cross section of
superradiant modes  $\sigma_{\ell m}(\omega)=-\Gamma_{\ell
m}(\omega)/{\mathcal F}_{\rm in}<0$.

The absorption cross section determines the ratio between the
outgoing and ingoing fluxes as \beq\label{finfout} \frac{{\mathcal
F}_{\rm out}}{{\mathcal F}_{\rm in}} =1-\frac{\omega^3}{(\ell+1)^2
4\pi}\sigma_{\ell m}\, \eeq (this is the relation in five
dimensions, see \cite{Gubser:1997qr} for generic dimension).
Superradiant modes, and only them, have $\sigma_{\ell m}<0$, and
therefore yield ${\mathcal F}_{\rm out}>{\mathcal F}_{\rm in}$, as
desired.

This argument shows that the extremal rotating system that we study
exhibits classical stimulated amplification for those modes that it
can spontaneously decay into, \ie modes that satisfy
\reef{sradbound4}. What happens away from extremality? In this case,
the system can spontaneously emit modes of any frequency
by the microscopic dual of Hawking radiation. Why, then, is there
superradiant amplification only for modes that satisfy
\reef{sradbound4}? The reason is known: the first law, applied to an
emission process from the black hole with $\delta E=-\omega$ and
$\delta J=-m$, states that \beq\label{area} \frac{\kappa}{8\pi
G}\delta \mathcal{A}_H =-(\omega -m\Omega_H)\,. \eeq Then, the
classical stimulated emission of a mode with $\omega > m\Omega_H$
would violate the area law $\delta \mathcal{A}_H\geq 0$
\cite{Bekenstein:1973mi}. So, classically, the emission of such
non-superradiant modes is strictly forbidden, while microscopically
it is allowed but statistically suppressed by a factor \beq
e^{\delta S}=e^{-(\omega-m\Omega_H)/T_H}\,. \eeq This is of course
the Boltzmann factor for Hawking radiation.

Sometimes the existence of the superradiant frequency bound
\reef{sradbound} is presented as a consequence of the area law. But
we see that the latter is important only in constraining the
classical, macroscopic process. Entropic considerations did not play
any role in our microscopic analysis, which nevertheless shows that
the superradiant bound on modes holds strictly at the microscopic
level for emission at zero temperature.

\setcounter{equation}{0}\section{\label{sec:Sugra}Emission rates: supergravity analysis}

The preceding analysis has provided a qualitative origin of the
superradiant ergoregion in rotating black holes at zero temperature.
We have also given a quantitative elementary derivation of the
superradiant frequency bound. A more precise match between the two
descriptions is obtained when one considers the actual emission
rates.

To do so, in this section we carry out the supergravity computation
of absorption cross sections and Hawking and superradiant emission
rates for a minimal scalar. We consider the most general case where
the black hole has all charges and rotations turned on, and the
scalar has generic quantum numbers for the frequency, spins, and
linear momentum along the $S^1$ string direction. At the end of the
section we particularize to the ergo-cold black hole in order to
isolate the effects of the ergosphere.

\subsection{\label{sec:properties metric}The D1-D5-P family of black
holes}

The D1-D5-P black hole solutions belong to type IIB supergravity
compactified to five dimensions on $T^4\times S^1$. The $T^4$ is
assumed to be much smaller than the $S^1$ so we view the system as a
a six-dimensional black string. The most general solution is
described by eight parameters: a parameter $M_0$ that measures
deviation away from supersymmetry; two spin parameters for rotation
in two orthogonal planes, $a_1,a_2$; three `boost' parameters,
$\delta_1,\delta_5,\delta_p$, which fix the D1-brane, D5-brane and
momentum charges, respectively; and two moduli: the radius $R$ of
the $S^1$, and the volume $V$ of the $T^4$. We choose units such
that the five-dimensional Newton constant is $G_5=G_{10}/2\pi
RV\equiv\pi/4$.

The metric of the six-dimensional black string is
\cite{cy,Cvetic:1998xh,Giusto:2004id}
\begin{eqnarray} \label{3charge}
ds^2&=&-\frac{f}{\sqrt{H_{1} H_{5}}}( dt^2 - dy^2)
+\frac{M_0}{\sqrt{H_{1} H_{5}}} (s_p dy - c_p dt)^2 +\sqrt{H_{1}
H_{5}} \left(\frac{ r^2 dr^2}{
(r^2+a_{1}^2)(r^2+a_2^2) - M_0r^2} +d\theta^2 \right)\nonumber \\
&&+\left( \sqrt{H_{1} H_{5}} - (a_2^2-a_1^2) \frac{( H_{1} + H_{5}
-f) \cos^2\theta}{\sqrt{H_{1}
H_{5}}} \right) \cos^2 \theta d \psi^2\nonumber \\
&&
 +\left(
\sqrt{H_{1} H_{5}} + (a_2^2-a_1^2) \frac{(H_{1} + H_{5} -f)
\sin^2\theta}{\sqrt{H_{1} H_{5}}}\right) \sin^2 \theta d \phi^2
  - \frac{M_0}{\sqrt{H_{1}
H_{5}}}(a_1 \cos^2 \theta d \psi + a_2 \sin^2 \theta d
\phi)^2\nonumber \\
&&
 - \frac{2M_0 \cos^2 \theta}{\sqrt{H_{1}
H_{5}}}{\bigl [}(a_1 c_1 c_5 c_p -a_2 s_1 s_5 s_p) dt + (a_2 s_1
s_5 c_p - a_1 c_1 c_5 s_p) dy {\bigr ]} d\psi \nonumber \\
&&
 -\frac{2M_0 \sin^2 \theta}{\sqrt{H_{1} H_{5}}}
{\bigl [} (a_2 c_1 c_5 c_p - a_1 s_1 s_5 s_p) dt + (a_1 s_1 s_5 c_p
- a_2 c_1 c_5 s_p) dy{\bigr ]} d\phi
  \,,
\end{eqnarray}
where we use the notation $c_i \equiv \cosh \delta_i$, $s_i \equiv
\sinh \delta_i$, and
\bea f(r)&=&r^2+a_1^2\sin^2\theta+a_2^2\cos^2\theta \,, \qquad
H_i(r)=f(r)+M_0 s_i^2\,, \:\,{\rm with}\:\, i=1,5 \,,\nonumber \\
g(r)&=&(r^2+a_1^2)(r^2+a_2^2)-M_0 r^2\,.
 \label{def f H}
 \eea
The dilaton and 2-form RR gauge potential will not be needed and can
be found in \cite{Giusto:2004id}.
%
%
We assume without loss of generality\footnote{The simultaneous
exchange $a_1\rightarrow -a_1$, $\delta_p\rightarrow -\delta_p$,
$y\rightarrow -y$ and $\psi\rightarrow -\psi$ is a symmetry of the
solution. The same is true for $a_2\rightarrow -a_2$,
$\delta_p\rightarrow -\delta_p$, $y\rightarrow -y$ and
$\phi\rightarrow -\phi$. So the solutions with $a_1a_2\leq 0$ are
physically equivalent to the solutions with $a_1a_2\ge 0$. For
definiteness we assume the latter.} \be a_1 \ge a_2\ge 0\,. \ee

Depending on the value of the parameters, the geometry can describe
a black hole, a naked singularity, a smooth soliton or a conical
singularity \cite{ross}. The black hole family of solutions is
described by the range $M_0\geq (a_1+a_2)^2$ and has horizons at
$g(r)=0$,
\begin{eqnarray}
 r_{\pm}^2= \frac{1}{2}\,(M_0-a_1^2-a_2^2) \pm \frac{1}{2}
 \sqrt{(M_0-a_1^2-a_2^2)^2-4a_1^2 a_2^2}\,. \label{r+-}
\end{eqnarray}

We are particularly interested in the existence of an ergoregion,
whose properties were discussed in \cite{ross}. The norm of the
Killing vector $\partial_t$,
 \beq  
|\partial_t|^2=-\frac{f-M_0 c_p^2}{\sqrt{H_1
 H_5}}\,,
\eeq becomes spacelike for $f(r)<M_0 c_p^2$. This defines a
six-dimensional ergoregion, which includes not only the effects of
rotation but also of the linear motion of the string. As we
mentioned above, and will prove below, the latter does not actually
contribute to superradiance. It is therefore more convenient to
consider the vector $\zeta=\partial_t+\tanh\delta_p \partial_y$ such
that, upon dimensional reduction (so linear momentum becomes
charge), its orbits define static asymptotic observers in the
five-dimensional black hole geometry, and whose causal character is
therefore associated to the rotation ergosphere. Specifically, its
norm
\begin{eqnarray}
 |\zeta|^2=-\frac{f-M_0}{\sqrt{H_1
 H_5}}\,,
 \label{ergoregionV}
\end{eqnarray}
becomes spacelike for $f(r)<M_0$ so a rotational ergosphere appears
at $f(r)=M_0$.

The ADM mass $M$, the angular momenta $(J_{\phi},J_{\psi})$ and the
gauge charges $(Q_1, Q_5, Q_p)$ are
\bea M &=& \frac{M_0}{2} \left[ \cosh (2\delta_1)
+ \cosh (2\delta_5) + \cosh (2 \delta_p)\right], \nonumber \\
J_{\phi} &=&   M_0 (a_2 c_1 c_5 c_p - a_1 s_1 s_5 s_p)\,,\nonumber \\
J_{\psi} &=&   M_0 (a_1 c_1 c_5 c_p - a_2 s_1 s_5 s_p)\,,\nonumber \\
 Q_i &=& M_0 s_i c_i\,, \qquad i=1,5,p \,. \label{ADMcharges}
\eea
The horizon angular velocities $\Omega_{\phi,\psi}$ along the Cartan
angles of $SO(4)$, $\phi$ and $\psi$, are more conveniently written
in terms of the Euler left and right rotations in $U(1)_L\times
U(1)_R \subset SU(2)_L\times SU(2)_R \simeq SO(4)$,
\bea 
\Omega_{\phi,\psi} = \frac{1}{2}\left(\Omega_R\pm\Omega_L
\right) \,, \qquad 
\Omega_{R,L}=\frac{2\pi}{\beta_H} \frac{a_2 \pm a_1}
 {\left[ M_0-(a_2\pm a_1)^2 \right]^{1/2}}\,. \label{omegaLR} \eea
Following \cite{Cvetic:1997uw}, from the surface gravities of the
inner and outer horizons $\kappa_\pm$ we introduce the temperatures
$\beta_{L,R}=1/T_{L,R}$
\bea 
\beta_{R,L} = \frac{ 2\pi}{\kappa_+} \pm \frac{2\pi}{\kappa_-} \,, \qquad 
\frac{1}{\kappa_{\pm}}=\frac{M_0}{2}\left[
 \frac{c_1c_5c_p + s_1s_5s_p} {\left[ M_0-(a_2+a_1)^2 \right]^{1/2} }
 \pm \frac{c_1c_5c_p - s_1s_5s_p}
                 {\left[ M_0-(a_2-a_1)^2 \right]^{1/2} }\right]\,. \label{betaLR}
\eea Observe that the Hawking temperature of the outer horizon is
related to $T_{L,R}$ as in \reef{TH}. Similarly, from the areas of
the inner and outer horizons we introduce $S_{L,R}$ such that
\bea 
S=S_L+S_R\,, \qquad 
 S_{R,L} = \pi M_0 \,\left(c_1c_5c_p
\mp s_1s_5s_p\right)\left[ M_0-(a_2\pm a_1)^2 \right]^{1/2}  \,.
\label{entropy} \eea
The horizon of the black string is also moving relative to
asymptotic observers that follow orbits of $\partial_t$. We can
compute the linear velocities for both the inner and outer horizons
\beq V_\pm=\frac{\pi M_0}{\beta_H}\left[
 \frac{c_1c_5s_p + s_1s_5c_p} {\left[ M_0-(a_2+a_1)^2 \right]^{1/2} }
 \pm \frac{c_1c_5s_p - s_1s_5c_p}
 {\left[ M_0-(a_2-a_1)^2 \right]^{1/2} }\right]\,,
\eeq
and introduce
 \beq \label{V(LR)}
V_{R,L}=-\frac{\beta_H}{\beta_{R,L}}(V_+\pm V_-)\,.
 \eeq
In terms of these, the velocity of the outer horizon, $V_+$, which
we also denote as $V_H$, is \beq\label{V_H}
 V_H=-\frac{T_H}{2}\left( \frac{V_L}{T_L} +\frac{V_R}{T_R} \right) \,,
\qquad \eeq These velocities become much simpler in the decoupling
limit where the D1 and D5 boosts are very large so the system is
near-supersymmetric, the numbers of anti-D1 and anti-D5 branes are
suppressed, and we can make contact with the dual CFT. In this
regime we approximate $c_{1,5}\simeq s_{1,5}\simeq
e^{\delta_{1,5}}/2$ and we find that
 \beq\label{decV} V_{L,R}\to \pm 1 \,,
 \eeq
which is microscopically interpreted as the fact that the momentum
excitations are chiral and massless\footnote{We are taking left
velocities and momenta as positive.}. Observe that in this regime we
recover eq.~\reef{VH}, which we had derived from the microscopic
two-sector system. The role that the inner horizon plays in defining
the microscopic magnitudes associated to the two chiral sectors,
emphasized in \cite{Cvetic:1997uw}, is very intriguing and not well
understood.

During the remainder of this section we will not need to restrict
ourselves to this near-supersymmetric regime. But our main interest
lies in extremal rotating black hole solutions. These correspond to
degenerate horizons, which appear when the two roots $r_\pm$
coincide. From \reef{r+-} we identify two possibilities:
\begin{itemize}
\item  The BPS black hole.

Obtained by taking the limit $M_0\to 0$, $a_{1,2}\to 0$, keeping the
mass, angular momenta and charges finite,
%
%
%
%
which requires $\delta_{1,5,p} \rightarrow \infty$. In this limit
\beq T_R\to 0\,,\quad T_L\neq 0\,,\qquad S_R\to 0\,,\quad S_L\neq
0\,,\qquad \Omega_{L,R}\rightarrow 0\,,\qquad -V_R,V_L\to V_H
\rightarrow 1\,. \eeq Also, $J_\phi+J_\psi\to 0$, the BPS bound is
saturated, the solution is supersymmetric, and the timelike Killing
vector that becomes null at the horizon is globally defined, so
there is no ergoregion. This is also clear from \reef{ergoregionV}.
This is the BMPV black hole.

\item  The ergo-cold black hole.

Obtained in the limit
\bea M_0 \rightarrow (a_1+a_2)^2\,,
 \label{ExtLim}
\eea
in which $T_H\to 0$ but now keeping $\Omega_R\neq 0$. Since $M_0\neq
0$ the BPS bound is not saturated and supersymmetry is absent. In
this limit, \bea T_R\to 0\,,\quad T_L\neq 0\,,\qquad S_R\to
0\,,\quad S_L\neq 0\,, \qquad \Omega_L\to 0\,,\quad \Omega_R\neq
0\,,\qquad -V_R \to V_H \,, \eea while $V_L$ takes no particular
value (unless we take the decoupling limit) and the conserved
charges $M$, $Q_i$ and $J_{\psi,\phi}$ are unconstrained other than
by the extremality condition. The horizon does rotate relative to
asymptotic observers, and there is an ergosphere, determined by
$f(r)=(a_1+a_2)^2$; see \reef{ergoregionV}. Observe that in contrast
to the BMPV solution, $J_\phi$ and $J_\psi$ are independent of each
other.
\end{itemize}
The BMPV black hole has been thoroughly studied, and it will only
serve us to emphasize the differences with the ergo-cold black hole,
which is our system of choice for the study of superradiance.

\subsection{\label{sec:Decay:Sugra}Absorption cross section and emission rate}

We consider a minimal scalar field, typically a graviton with
polarization in the internal $T^4$ in the compactification of the
IIB theory to six dimensions. The field satisfies the massless
Klein-Gordon equation in the general three-charge black string
geometry,
\begin{eqnarray}
\partial_{\mu}
\left(\sqrt{-g}\,g^{\mu\nu} \partial_{\nu}\Phi \right)=0\,.
 \label{klein}
\end{eqnarray}
Introducing the ansatz
\begin{eqnarray}
\Phi=\exp\left[-i\omega t+ip y+i m_{\psi} \psi +i m_{\phi} \phi
\right]\, \chi(\theta)\,h(r) \,,
 \label{separation ansatz}
\end{eqnarray}
and the separation constant $\Lambda$, the wave equation can be
separated. The angular equation is
\begin{eqnarray}
 \frac{1}{\sin{2\theta}}\frac{d}{d\theta}\left
(\sin{2\theta}\,\frac{d\chi}{d\theta}\right )+{\biggr
[}\Lambda-\frac{m_{\psi}^2}{\cos^2\theta}-\frac{m_{\phi}^2}{\sin
^2\theta}+(\omega ^2-p ^2)(a_1 ^2\sin ^2\theta+a_2^2\cos^2\theta)
{\biggr ]}\chi=0 \label{angeq}\, .
\end{eqnarray}
This angular equation (plus regularity requirements) is a
Sturm-Liouville problem, and the solutions are higher-dimensional
spin-weighted spheroidal harmonics. We can label the corresponding
eigenvalues $\Lambda$ with an index $\ell$,
$\Lambda(\omega)=\Lambda_{\ell} (\omega)$ and therefore the
wavefunctions form a complete set over the integer $\ell$. In the
general case, the problem consists of two coupled second order
differential equations: given some boundary conditions, one has to
compute simultaneously both values of $\omega$ and $\Lambda$ that
satisfy these boundary conditions. However, for vanishing $a_i ^2$
we get the (five-dimensional) flat space result, $\Lambda=\ell
(\ell+2)$, and the associated angular functions are given by Jacobi
polynomials. For non-zero, but  small $(\omega ^2-p ^2)a_i ^2$ we
have
\begin{eqnarray} \Lambda=\ell (\ell+2)+\mathcal{O}
\left (a_i^2(\omega^2-p ^2)\right) \label{app} \,.
\end{eqnarray}
The integer $\ell$ is constrained to be $\ell \geq
|m_{\phi}|+|m_{\psi}|$, and can only take even (odd) values when
$|m_{\phi}|+|m_{\psi}|$ is even (odd) \cite{Berti:2005gp}---this
follows from the fact that the scalar $\ell^\mathrm{th}$ wave is in
the $(\ell/2,\ell/2)$ of $SU(2)_R\times SU(2)_L$. The angular
coordinates $\phi,\psi$ are periodic with period $2\pi$ so
$m_{\phi}$, $m_{\psi}$ must take integer values. Our waves have
positive frequency $\omega>0$.

The radial wave equation can be written in a form that is
particularly appropriate to find its solutions. Introduce the new
radial coordinate
\begin{eqnarray}
 x=\frac{r^2-\frac{1}{2}(r_+^2+r_-^2)}{r_+^2-r_-^2}\,,
 \label{def:x}
\end{eqnarray}
which maps $r=(r_-,r_+,\infty)\leftrightarrow x=(-1/2,1/2,\infty)$.
Introduce also
\begin{eqnarray}
 m_{L,R}=\frac{1}{2}\left(m_{\phi}\mp m_{\psi}\right)\,.
 \label{def:mLR}
\end{eqnarray}
The radial wave equation is then
\begin{eqnarray}
&&\hspace{-2cm}\partial_x\left[\left(x-\frac{1}{2}\right)\left(x+\frac{1}{2}\right)\partial_x
h\right] +\frac{1}{4}\left[  (\omega^2 - p^2)
\left(r_+^2-r_-^2\right)x-(\Lambda-U) \right]h  \nonumber\\
&& \hspace{6cm}+\frac{1}{4}\left[
\frac{\Sigma_+^{\,2}}{\left(x-\frac{1}{2}\right)}
  - \frac{\Sigma_-^{\,2}}{\left(x+\frac{1}{2}\right)}
  \right]h=0\,,
 \label{radialeq}
\end{eqnarray}
where we defined
\begin{eqnarray}
&& \Sigma_{\pm}=\frac{\omega}{\kappa_{\pm}} \mp
m_L\frac{\Omega_L}{\kappa_+}-m_R\frac{\Omega_R}{\kappa_+}- p\frac{
V_{\pm}}{\kappa_+}\,, \nonumber \\
 &&
U=(\omega^2 - p^2) \left[ \frac{1}{2}(r_+^2+r_-^2)+ M_0 (s_1^2 +
s_5^2)\right] + (\omega c_p + p s_p)^2 M_0\,.
 \label{def:U}
\end{eqnarray}
Equation \reef{radialeq} was first written (though in a much less
compact form) in \cite{Cardoso:2007ws}. For $p=0$ there is no
dynamics associated to the sixth direction and (\ref{radialeq})
reduces to the wave equation studied in \cite{Cvetic:1997uw} for the
scattering of a neutral scalar off the five-dimensional D1-D5-P
black hole.

\subsubsection{\label{sec:BH Near region}Near-region wave equation
and solution}
In the near-region, the term $p^2 \left(r_+^2-r_-^2\right)x$ is
suppressed and the radial wave equation reduces to
\begin{eqnarray}
\partial_x\left[\left(x-\frac{1}{2}\right)\left(x+\frac{1}{2}\right)\partial_x
h\right] + \frac{1}{4}\left[ -(\Lambda-U)+
\frac{\Sigma_+^{\,2}}{\left(x-\frac{1}{2}\right)}
  - \frac{\Sigma_-^{\,2}}{\left(x+\frac{1}{2}\right)}
  \right]h=0\,.
 \label{near wave eq}
\end{eqnarray}
To find the analytical solution of this equation, define the new
radial coordinate,
\begin{eqnarray}
z=x+\frac{1}{2}\, , \qquad r=(r_-,r_+,\infty)\leftrightarrow
x=(-1/2,1/2,\infty)\leftrightarrow z=(0,1,\infty)\,,
 \label{def:z}
\end{eqnarray}
and introduce the new wavefunction
\begin{eqnarray}
h=z^{-i \,\frac{1}{2}\Sigma_-} (z-1)^{-i \,\frac{1}{2}\Sigma_+} \,F
\,,
 \label{hypergeometric function}
\end{eqnarray}
The near-region radial wave equation can then be written as
\begin{eqnarray}
& &  \hspace{-1cm}
 z(1-z)\partial_z^2 F+ {\biggl [} (1-i\,\Sigma_-)-\left[ 2- i\,(\Sigma_+ +\Sigma_-) \right ]\,z {\biggr ]}
\partial_z F \nonumber \\
&& \hspace{2cm}+ \left [ i\,\frac{1}{4}(\Sigma_+ +\Sigma_-)[2-
i\,(\Sigma_+ +\Sigma_-)]+(\Lambda-U) \right ] F=0\,,
 \label{near wave hypergeometric}
\end{eqnarray}
which is a standard hypergeometric equation \cite{abramowitz},
$z(1\!-\!z)\partial_z^2 F+[c-(a+b+1)z]\partial_z F-ab F=0$, with
\begin{eqnarray}
& & \hspace{-0.5cm} a=\xi-\frac{i}{2}\,(\Sigma_+ +\Sigma_-) \,,
\qquad b=1-\xi-\frac{i}{2}\,(\Sigma_+ +\Sigma_-) \,, \qquad c=1-
i\,\Sigma_- \,,
 \label{hypergeometric parameters}
\end{eqnarray}
where we defined
\begin{eqnarray}
\xi=\frac{1}{2}\left(1+\sqrt{1+\Lambda-U} \right)\,.
 \label{xi}
\end{eqnarray}
Its most general solution in the neighborhood of $z=1$ (\ie $r=r_+$)
is $A_H^{in}\, z^{-b} F(b,b-c+1,a+b-c+1,\frac{z-1}{z})+A_H^{out}\,
z^{a-c}(z-1)^{c-a-b}F(c-a,1-a,c-a-b+1,\frac{z-1}{z})$. Using
(\ref{hypergeometric function}), one finds that the solution of the
near-region equation is
\begin{eqnarray}
h &=& A_H^{in}\,
\left(x-\frac{1}{2}\right)^{-i\,\frac{1}{2}\Sigma_+}\,\left(x+\frac{1}{2}\right)^{-\xi+i\,\frac{1}{2}\Sigma_+
}\,
F\left(b,b-c+1,a+b-c+1,\frac{x-\frac{1}{2}}{x+\frac{1}{2}}\right)\nonumber \\
& &
+A_H^{out}\,\left(x-\frac{1}{2}\right)^{+i\,\frac{1}{2}\Sigma_+}\,\left(x+\frac{1}{2}\right)^{-\xi}\,
F\left(c-a,1-a,c-a-b+1,\frac{x-\frac{1}{2}}{x+\frac{1}{2}}\right)
\,.
 \label{hypergeometric solution}
\end{eqnarray}
The first term represents an ingoing wave at the horizon
$x=\frac{1}{2}$, while the second term represents an outgoing wave
at the horizon. The computation of the absorption cross-section is a
classical problem where outgoing waves at the horizon are forbidden,
so we set $A_H^{out}=0$. Furthermore, we need the large $r$,
$x\rightarrow \infty$ behavior of the ingoing near-region solution.
We use the $z \rightarrow 1-z$ transformation law for the
hypergeometric function \cite{abramowitz},
\begin{eqnarray}
&
\hspace{-2cm}F\left(b,b-c+1,a+b-c+1,\frac{x-\frac{1}{2}}{x+\frac{1}{2}}\right)=
\frac{\Gamma(a+b-c+1)\Gamma(a-b)}{\Gamma(a-c+1)\Gamma(a)}
 \,F\left(b,b-c+1,-a+b+1,\frac{1}{x+\frac{1}{2}}\right) & \nonumber \\
& \hspace{4cm} +\left(x+\frac{1}{2}\right)^{a-b}
\frac{\Gamma(a+b-c+1)\Gamma(-a+b)}{\Gamma(b)\Gamma(b-c+1)}
 \,F\left(a-c+1,a,a-b+1,\frac{1}{x+\frac{1}{2}}\right), & \nonumber \\
&
 \label{transformation law}
\end{eqnarray}
the property $F(a,b,c,0)=1$, and $x\pm \frac{1}{2}\sim x$. The large
$x$ behavior of the ingoing near-horizon solution is then
\begin{eqnarray}
h &\sim& A_H^{in}{\biggl [}
 \frac{\Gamma\left[1-i\Sigma_+\right] \,\Gamma\left[1-2\xi\right]}
{\Gamma\left[1-\xi-i\,\frac{1}{2}(\Sigma_+-\Sigma_-)\right]
\Gamma\left[1-\xi-i\,\frac{1}{2}(\Sigma_+ +\Sigma_-)\right]}\:
x^{-\xi}\nonumber \\
& &
 \qquad\qquad +\frac{\Gamma\left[1-i\Sigma_+\right] \,\Gamma\left[1-2\xi\right]}
{\Gamma\left[\xi-i\,\frac{1}{2}(\Sigma_+-\Sigma_-)\right]
\Gamma\left[\xi-i\,\frac{1}{2}(\Sigma_+ +\Sigma_-)\right]}\:
x^{\xi-1} {\biggr ]}.
 \label{near field-large r}
\end{eqnarray}
\subsubsection{\label{sec:BH Far region}Far-region wave equation and
solution}

In the far-region, the terms $\left( x\pm \frac{1}{2} \right)^{-1}$
are suppressed, and $x\pm \frac{1}{2}\sim x$. The radial wave
equation can be written as
\begin{eqnarray}
\partial_x^2 (xh)+ \left [ \frac{(\omega^2 - p^2)(r_+^2-r_-^2)}{4x}
-\frac{\Lambda-U}{4x^2} \right ] (xh)=0\,.
 \label{far wave eq}
\end{eqnarray}
The most general solution of this equation is a linear combination
of Bessel functions \cite{abramowitz},
\begin{eqnarray}
h=x^{-1/2}\left [ A_{\infty}^+ J_{\,2\xi-1}(\mu x^{1/2})+
A_{\infty}^- J_{\,1-2\xi}(\mu x^{1/2})\right ]\,,
 \label{far field}
\end{eqnarray}
where $\xi$ was defined in (\ref{xi}) and
\begin{eqnarray}
\mu=\left[(\omega^2 - p^2)(r_+^2-r_-^2)\right]^{1/2}\,.
 \label{def:mu}
\end{eqnarray}
We want to study the scattering process so we require real $\mu$ \ie
$\omega>|p|$. Using the asymptotic properties of the Bessel function
\cite{abramowitz}, we find that for small $\mu x^{1/2}$ the
far-region solution has the behavior
\begin{eqnarray}
h \sim \,A_{\infty}^+\, \frac{(\mu/2)^{2\xi-1}}{\Gamma(2\xi)}\:
x^{\xi-1} + A_{\infty}^-\, \frac{(\mu/2)^{1-2\xi}}{\Gamma(2-2\xi)}\:
x^{-\xi}.
 \label{far field-small r}
\end{eqnarray}
while for large $\mu x^{1/2}$ it reduces to
\begin{eqnarray}
&& \hspace{-1cm} h \sim
\frac{1}{2}\sqrt{\frac{2}{\pi\mu}}x^{-3/4}{\biggl \{}
 \left[A_{\infty}^+
e^{-i\pi(-\xi+1/4)}+A_{\infty}^- e^{-i\pi(\xi-3/4)} \right] e^{-i\mu
\sqrt{x}} \nonumber \\
& &\hspace{3cm}+ \left[A_{\infty}^+ e^{i\pi(-\xi+1/4)}+A_{\infty}^-
e^{i\pi(\xi-3/4)} \right] e^{i\mu \sqrt{x}}{\biggr \}}.
 \label{far field-large r}
\end{eqnarray}
The first term represents an incoming wave while the second term
describes an outgoing solution.

\subsubsection{\label{sec:BH Matching}Matching the near-region and
the far-region solutions}

There is an intermediate region for $x$ where the approximations in
both the near and far regions can be simultaneously satisfied. In
this overlapping region we can match the large $x$ behavior of the
near-region solution to the small $x$ behavior of the far-region
solution. This allows to fix the amplitude ratios. Matching
(\ref{near field-large r}) with
 (\ref{far field-small r}) yields then
\begin{eqnarray}
& &
\frac{A_H^{in}}{A_{\infty}^+}=\left(\frac{\mu}{2}\right)^{2\xi-1}
\frac{\Gamma\left[\xi-i\,\frac{1}{2}(\Sigma_+-\Sigma_-)\right]
\Gamma\left[\xi-i\,\frac{1}{2}(\Sigma_+ +\Sigma_-)\right]}
 {\Gamma(2\xi)\Gamma(2\xi-1)\Gamma\left[1-i\Sigma_+\right]}
\,,\nonumber \\
& &
\frac{A_{\infty}^-}{A_{\infty}^+}=\left(\frac{\mu}{2}\right)^{2(2\xi-1)}
\frac{\Gamma(2-2\xi)\Gamma(1-2\xi)}{\Gamma(2\xi)\Gamma(2\xi-1)}
\frac{\Gamma\left[\xi-i\,\frac{1}{2}(\Sigma_+-\Sigma_-)\right]
\Gamma\left[\xi-i\,\frac{1}{2}(\Sigma_+ +\Sigma_-)\right]}
 {\Gamma\left[1-\xi-i\,\frac{1}{2}(\Sigma_+-\Sigma_-)\right]
\Gamma\left[1-\xi-i\,\frac{1}{2}(\Sigma_+ +\Sigma_-)\right]}\,.\nonumber \\
& &
 \label{relation beta-alpha}
\end{eqnarray}
The first relation will be needed to compute the absorption cross
section. In the second relation we note the presence of the factor
$\mu^{2\xi-1}$, where $\mu$ is defined in $(\ref{def:mu})$. We want
$\xi \in \mathbb{R}$ which implies $2\xi-1>0$. Therefore, for
$\mu\ll 1$, \ie for low frequency scattering or for
near-supersymmetric solutions (decoupling limit), one has
$|A_{\infty}^-|\ll |A_{\infty}^+|$. This regime allows to
considerably simplify (\ref{far field-large r}).

\subsubsection{\label{sec:rate}Absorption cross-section, Hawking and superradiant emission rate}

The radial flux associated with our radial wave equation is
\begin{eqnarray}
\mathcal{F}=\frac{1}{2i}\left(h^*\frac{g(r)}{r}\partial_r h- h
\frac{g(r)}{r}\partial_r h^*\right).
 \label{flux}
\end{eqnarray}
The incoming flux from infinity $\mathcal{F}_{in}$ is computed using
(\ref{far field-large r}). Near the decoupling regime
$|A_{\infty}^-|\ll
 |A_{\infty}^+|$, this yields
\begin{eqnarray}
\mathcal{F}_{in}=-\frac{r_+^2-r_-^2}{2\pi}\left|A_{\infty}^+\right|^2,
 \label{incflux}
\end{eqnarray}
where the minus sign signals incoming flux. On the other hand, use
of the ingoing contribution of (\ref{hypergeometric solution})
yields for the absorbed flux at the horizon,
\begin{eqnarray}
\mathcal{F}_{abs}=-\Sigma_+(r_+^2-r_-^2)\left|A_H^{in} \right|^2.
 \label{absflux}
\end{eqnarray}
The absorption probability is the ratio of the above fluxes,
\begin{eqnarray}
1-|S_{\ell}|^2=\frac{\mathcal{F}_{abs}}{\mathcal{F}_{in}}\,,
 \label{probability}
\end{eqnarray}
and the absorption cross-section of the $\ell^{\rm th}$ partial wave
is
\begin{eqnarray}
\sigma_{\ell,p,m_{R,L}} =\frac{4\pi}{\omega^3}(\ell
+1)^2\left(1-|S_{\ell}|^2\right)\,.
 \label{crosssection}
\end{eqnarray}
In general, the factor multiplying the absorption probability
depends on the spacetime dimension through the codimension of the
absorbing object (see, \eg \cite{Gubser:1997qr}). So for a
six-dimensional black string we use the same factor as for a
five-dimensional black hole. Collecting the results, the absorption
cross-section is
\begin{eqnarray}
\sigma_{\ell,p,m_{R,L}}=\frac{4\pi (\ell +1)^2}{\omega^3}\,\beta_H
\varpi \left[
\frac{1}{4}(\omega^2-p^2)(r_+^2-r_-^2)\right]^{2\xi-1}\left|
\frac{\Gamma\left(\xi-i\frac{\beta_L\varpi_L}{2\pi}\right)\Gamma\left(\xi-i\frac{\beta_R
\varpi_R}{2\pi}\right)}{\Gamma\left(2\xi\right)\Gamma\left(2\xi-1\right)
\Gamma\left(1-i\frac{\beta_H \varpi}{2\pi}\right)} \right|^2,
 \label{crosssectionend}
\end{eqnarray}
where we defined
\begin{eqnarray}
\varpi=\omega-p\,{V_H}-m_L\Omega_{L}-m_R\Omega_{R}
 \,, \qquad \qquad
\varpi_{L,R}=\frac{1}{2}\left(\omega+pV_{L,R}\,\right)-m_{L,R}\Omega_{L,R}\frac{\beta_H}{\beta_{L,R}}
\,.
 \label{def:Sigmas}
\end{eqnarray}
Observe in the latter equation the presence of
$\Omega_{L,R}\frac{\beta_H}{\beta_{L,R}}$, which correspond to the
chemical potentials $\mu_{L,R}$ of the microscopic two-sector system
\reef{Omu}.

The matching \reef{relation beta-alpha} was performed in the low
frequency regime of waves with wavelength much larger than the
typical size of the black hole. This is actually the regime of
relevance when comparing to the microscopic dual, in which the
excitations near the horizon are (almost) decoupled from the
asymptotic region, and we only allow a little leakage of energy
between the two regions. The latter corresponds to coupling the dual
CFT to a bulk scalar. Using (\ref{app}) and (\ref{def:U}) this is
the range of parameters where
\begin{eqnarray}
 U\ll \Lambda \simeq \ell (\ell+2) \quad \Rightarrow \quad \xi \simeq \frac{\ell+2}{2}\,.
 \label{restriction}
\end{eqnarray}
In particular, since $\xi$ is integer or half-integer, the following relations
\cite{abramowitz}
\begin{eqnarray}
 && \left|\Gamma(n-i z)\right|^2=\Gamma(n-i z)\Gamma(n+i z)\,, \qquad
 \Gamma(n\pm i z)= \Gamma(1\pm i z) \prod_{j=1}^{n-1}\left( j^2+z^2
 \right)\,,
\nonumber \\
 && |\Gamma(1-i z)|^2=\frac{\pi z}{\sinh(\pi z)}\,, \qquad
 \left|\Gamma\left(\frac{1}{2}-i
 z\right)\right|^2=\frac{\pi}{\cosh(\pi z)}
 \,,
 \label{gammaprop}
\end{eqnarray}
are useful.
Thus we can rewrite \reef{crosssectionend}
as
\begin{eqnarray}
\sigma_{\ell,p,m_{R,L}}&=& \frac{8\pi}{(\ell!)^4} \frac{1}{\omega^3}\left(
(\omega^2-p^2)\frac{\mathcal{A}_H^{(5)}}{4\pi \beta_H}\right)^{\ell+1}
 \sinh\left(\frac{\beta_H\varpi}{2}\right)\nonumber\\
&&     \times \left| \Gamma \left( \frac{\ell+2}{2} + i
{\beta_L\varpi_L \over 2 \pi} \right)
      \Gamma \left( \frac{\ell+2}{2} + i {\beta_R \varpi_R \over 2 \pi} \right)
      \right|^2\!,
 \label{absorptionSUGRA}
\end{eqnarray}
where we have used $r_+^2-r_-^2=\mathcal{A}_H^{(5)}/(4 G_5\beta_H)$
with $\mathcal{A}_H^{(5)}$ the area of the five-dimensional black
 hole, and in our units $G_5=\pi/4$.

By detailed balance, the decay rate is the absorption cross-section
divided by the thermal Bose-Einstein occupation number,
\begin{eqnarray}
\Gamma_{\ell,p,m_{R,L}}=\frac{\sigma_{\ell,p,m_{R,L}}}{e^{\beta_H
\varpi}-1} \,.
 \label{decayrate}
\end{eqnarray}

Use of \reef{gammaprop} also allows to write the decay rate in terms
of thermal factors. We have to distinguish the cases of even and odd
angular quantum number $\ell$. For even $\ell$,
(\ref{crosssectionend}), (\ref{decayrate}), (\ref{gammaprop}) give
the decay rate,
\begin{eqnarray}
{\rm Even}\:\:\ell: \qquad \Gamma_{\ell,p,m_{R,L}}&=&
\frac{4\pi}{(\ell!)^4} \left[
(\omega^2-p^2)\frac{\mathcal{A}_H^{(5)}}
 {4\pi}\right]^{\ell+1}
\frac{\varpi_L \varpi_R}{\omega^3}\,\frac{\beta_L \beta_R}{\beta_H}
\left(e^{\beta_L \varpi_L}-1\right)^{-1}
\left(e^{\beta_R \varpi_R}-1\right)^{-1} \nonumber \\
&& \times \prod_{j=1}^{\ell/2}\left[
j^2+\left(\frac{\beta_L\varpi_L}{2\pi}\right)^2\right] \left[
\left(\frac{j}{\beta_H}\right)^2+\left(\frac{\beta_R\varpi_R}{2\pi\beta_H}\right)^2\right].
 \label{rate:evenl}
\end{eqnarray}
For odd $\ell$, the decay rate is
\begin{eqnarray}
&& \hspace{-2cm}{\rm Odd}\:\:\ell: \qquad \Gamma_{\ell,p,m_{R,L}} =
\frac{2(2\pi)^3}{(\ell!)^4} \left[
(\omega^2-p^2)\frac{\mathcal{A}_H^{(5)}}
 {4\pi}\right]^{\ell+1}
\frac{1}{\omega^3} \left(e^{\beta_L \varpi_L}+1\right)^{-1}
\left(e^{\beta_R \varpi_R}+1\right)^{-1} \nonumber \\
&&\hspace{2.2cm} \times \prod_{j=1}^{(\ell+1)/2}\left[
\left(j-\frac{1}{2}\right)^2+\left(\frac{\beta_L\varpi_L}{2\pi}\right)^2\right]
\left[
\left(\frac{2j-1}{2\beta_H}\right)^2+\left(\frac{\beta_R\varpi_R}{2\pi\beta_H}\right)^2\right].
 \label{rate:oddl}
\end{eqnarray}
As observed in \cite{Maldacena:1997ih}, for even $\ell$ there appear
left and right bosonic thermal factors (\ref{rate:evenl}) while for
odd $\ell$ they are fermionic thermal factors. This is already a
hint of the microscopic degrees of freedom responsible for the
radiation---taking into account that the bosonic factors can emerge
as effective ones from even numbers of fermions
\cite{Callan:1996tv,Mathur:1997et}.

\subsection{Superradiant emission rate from the ergo-cold black hole}

These emission rates contain effects of Hawking radiation as well as
superradiance. As explained in the introduction, in order to
eliminate the former we take an extremal, zero temperature limit,
while at the same time we want to preserve the superradiant
ergoregion.

In the case of the supersymmetric BMPV black hole, neither thermal
nor superradiant emission are present. In the limit to this solution
\bea
 \lim_{\beta_R\rightarrow\infty}
\varpi_R=\frac{\omega}{2}>0\,,
 \label{BPS:limSigma} \eea
and the positivity of $\varpi_R$ implies that, in \reef{rate:evenl} and
\reef{rate:oddl}, the right thermal
factor $(e^{\beta_R \varpi_R}\pm 1)^{-1}\to 0$, so $\Gamma_{\ell,p,
m_{R,L}}=0$. This is as it should be, since this a BPS state. The
absorption cross section is positive for any quantum numbers of the
wave, so stimulated emission cannot occur either.

The ergo-cold black hole is obtained in the limit in which
$\beta_R\to\infty$ while $\Omega_R$ remains finite. In this case
\bea
 \lim_{\beta_R\rightarrow\infty}
 \varpi_L=\frac{1}{2}\left(\omega+
p V_L\right)-m_L\,\frac{\pi(a_2-a_1)\sqrt{a_1a_2}}{\beta_L}
 \,, \qquad
\lim_{\beta_R\rightarrow\infty} \varpi_R=\frac{1}{2}\left(\omega-p
V_H - m_R\Omega_R\right) \,.
 \label{ExtLim:limSigma} \eea
Now $\varpi_R$ can take negative values, so the decay rates do not
vanish for all modes but contain a factor of a step function
\bea
 \lim_{\beta_R\rightarrow\infty}
 \left(e^{\beta_R \varpi_R}\pm 1\right)^{-1}=
\mp\Theta(-\varpi_R)\,,
 \label{ExtLim:limthermal}
\eea
so the emission decay rate is
\begin{eqnarray}
&& {\rm Even}\:\:\ell: \quad \Gamma_{\ell,p,m_{R,L}} =
\Theta(-\varpi_R) \frac{8\pi^2}{(\ell!)^4} \left[
(\omega^2-p^2)\frac{\mathcal{A}_H^{(5)}}
 {4\pi^2}\right]^{\ell+1}
\frac{\beta_L\varpi_L \left|\varpi_R \right|^{\ell+1}}
{\omega^3\left(e^{\beta_L \varpi_L }-1\right)}
\prod_{j=1}^{\ell/2}\left[
j^2+\left(\frac{\beta_L\varpi_L }{2\pi}\right)^2\right], \nonumber \\
&&{\rm Odd}\:\:\ell: \quad \Gamma_{\ell,p,m_{R,L}} =
\Theta(-\varpi_R)\frac{2(2\pi)^3}{(\ell!)^4} \left[
(\omega^2-p^2)\frac{\mathcal{A}_H^{(5)}}
 {4\pi^2 }\right]^{\ell+1}
\frac{ \left|\varpi_R \right|^{\ell+1}}{\omega^3\left(e^{\beta_L
\varpi_L }+1\right)}\nonumber \\
&&\hspace{7cm} \times
 \prod_{j=1}^{(\ell+1)/2}\left[
\left(j-\frac{1}{2}\right)^2+\left(\frac{\beta_L\varpi_L
}{2\pi}\right)^2\right].
 \label{ExtLim:limrate}
\end{eqnarray}

Thus we have derived the superradiant bound \reef{wp}. The ergo-cold
black hole can only emit modes that satisfy $\varpi_R<0$. The
absorption cross section is positive or negative depending on
whether $\varpi_R$ is positive or negative, so when $\varpi_R <0$,
and only then, superradiant amplification occurs.

We can also see that there cannot be any spinless, pure momentum
superradiance. An oscillating wave near infinity must have
$\omega>|p|$. Technically, this follows from the reality requirement
of quantities like (\ref{def:mu}) or (\ref{ExtLim:limrate}).
Physically, $\omega^2-p^2>0$ for a wave propagating in the
asymptotically flat region. According to (\ref{ExtLim:limSigma}),
spinless superradiant modes require $\omega< p V_H$. But
(\ref{V(LR)}) gives at extremality $V_H=\frac{c_1c_5s_p +
s_1s_5c_p}{c_1c_5c_p + s_1s_5s_p}$ so $|V_H|\leq 1$ and $|p V_H|\leq
|p|$. Then, none of these superradiant momentum modes can exist as
propagating waves at infinity: if emitted by the black hole, they
will be reflected back to it before getting to the asymptotic
region. This is a general feature present in black string
backgrounds \cite{Cardoso:2004zz,Dias:2006zv}.

\setcounter{equation}{0}\section{\label{sec:dualcft}Microscopic description}

\subsection{\label{sec:IdentifyCFTstates}The dual CFT state}

The CFT state dual to the ergo-cold black hole is most easily
identified by analyzing the solution in the decoupling limit. This
is a low energy limit, keeping the energies finite in string units,
which is obtained taking $\alpha^{\prime} \rightarrow 0$ and
$\delta_{1,5} \rightarrow \infty$ while keeping $r
(\alpha^{\prime})^{-1}$, $M (\alpha^{\prime})^{-2}$, $a_{1,2}
(\alpha^{\prime})^{-1}$, and $Q_{1,5} (\alpha^{\prime})^{-1}$ fixed.
For the general black hole geometry, this has been shown to result
in a twisted fibration of $S^3$ over the BTZ black hole
~\cite{Cvetic:1998xh}. The CFT states dual to the extremal black
holes we have been studying can be identified using the map
introduced in \cite{CFTmap}. This yields the R-charges $(j,\bar{j})$
and conformal dimensions $(h,\bar{h})$ of the CFT state in terms of
parameters of the supergravity solution. Introducing the AdS$_3$
curvature radius $\ell_3$, BTZ black hole mass $M_3$,
\begin{eqnarray}
&& \ell_3^2 = \sqrt{Q_1 Q_5}, \nonumber \\
 && M_3=\frac{R^2}{\ell^4} \left[ (M-a_1^2-a_2^2)
\left(c_p^2+s_p^2\right)
 + 4 a_1 a_2 s_pc_p \right],
 \label{paramDecoup}
\end{eqnarray}
and central charge $c=3\ell_3/2$, the following values are obtained
for the two extremal rotating black holes:

\begin{itemize}
\item BMPV black hole:
\begin{eqnarray}
&&j=\frac{c}{6}\frac{R}{\ell_3^4}\,J_{L}\,,\qquad h=
\frac{c}{24}\left(1+2M_3+\frac{4R^2}{\ell_3^8}\,J_{L}^2\right)\,, \nonumber\\
&& \bar{j}=0\,,\qquad \qquad
 \bar{h}=\frac{c}{24}\,.
 \label{cft:BPS}
\end{eqnarray}

\item Ergo-cold black hole:
\begin{eqnarray}
&&j= \frac{c}{6}\frac{R}{\ell_3^4}\,J_{L}\,,\qquad
h= \frac{c}{24}\left(1+2M_3+\frac{4R^2}{\ell_3^8}J_{L}^2\right)\,, \nonumber\\
&& \bar{j}=
  \frac{c}{6}\frac{R}{\ell_3^4}\,J_{R}\,,\qquad
\bar{h}=
  \frac{c}{24}\left(1+\frac{4R^2}{\ell_3^8}J_{R}^2\right)\,.
 \label{cft:ExtErgo}
\end{eqnarray}
\end{itemize}

To interpret these results we note that the conformal dimensions
receive contributions of three kinds, \beq h=h_0+l_0 + \frac{6
j^2}{c}\,,\qquad \bar h=\bar h_0+\bar l_0 + \frac{6 \bar j^2}{c}\,.
\eeq Here $(h_0,\bar h_0)=(c/24,c/24)$ correspond to the energy of
the Ramond ground-state. On top of this, the left sector has in both
cases an excitation energy given by the Virasoro level $l_0 =\ell_3
M_3/8$: this is the energy of its thermal excitations, which give
the system a Cardy entropy \beq S_L=2\pi\sqrt{c l_0/6}\,. \eeq
Additionally, the left sector contains some polarized fermions,
which yield a charge $j$. The Kac-Moody level of the
superconformal current algebra is $k=2c/3$. The Sugawara stress-energy tensor of
the $SU(2)$ current then yields an energy $(2j)^2/k=6 j^2/c$, where $2j$
appears since we are normalizing $j$ to be half-integer quantized.

The right sector in both black holes is at zero level, $\bar l_0=0$,
so they are at vanishing temperature. But there is a crucial
difference between the two states: whereas in the BMPV black hole
this sector is in a Ramond ground state, in the ergo-cold black
hole it is filled with polarized fermions, giving charge $\bar j$
and additional energy $6 \bar j^2/c$ that lifts the system above the
BPS state. This is the microscopic picture that we are advocating
for this black hole.

\subsection{\label{sec:Decay:micro}Emission rate and absorption cross section}

A coupling of the schematic form
\begin{eqnarray}
 \label{Sint}
S_{\rm int} \propto \int dtdx  \,
\partial^{\ell}\Phi(t,x,\vec{x}\!=\!0) \O(t,x) \,,
  \end{eqnarray}
($t,x$ are worldsheet coordinates and $\vec{x}$ are directions
transverse to the string) describes the interaction of the
$\ell^{\rm th}$ partial wave of the bulk scalar $\Phi$ with an
operator $\O(t,x)$ of the CFT of conformal dimension
$(1+\ell/2,1+\ell/2)$. We build the latter out of a pair of bosons
$\partial_\pm X$, and $\ell$ pairs of left and right fermions
$\psi_L \bar \psi_R$. This coupling gives a decay rate of the CFT
into a scalar mode with quantum numbers $\omega$, $\ell$, $p$, $m_{R,L}$,
of the form
\beq\label{decayG} \Gamma_{\ell,p,m_{R,L}}(\omega) =\,
\mathcal{V} \int dx^+ dx^- e^{-i\varpi_R x^- -i\varpi_L
x^+}{\mathcal G}(t-i\varepsilon,x)\,,
\eeq
where $x^{\pm}=t\pm
x$, the Green's function is
\beq\label{Gtx} {\mathcal G}(t,x)=\langle
\O^\dagger(t,x)\O(0)\rangle\,,
\eeq
with the $i\varepsilon$
prescription in \reef{decayG} corresponding to emission, $\mathcal{V}$ is a factor from
the interaction vertex to be discussed below, and
\beq
\varpi_{L,R}=\frac{1}{2}\left(\omega \pm p\right) -m_{L,R}\mu_{L,R}
\eeq
account for the presence of left and right sectors with chiral
momenta $(\omega\pm p)/2$ and chemical potentials $\mu_{L,R}$ for
the R-charges $m_{R,L}$, given by \reef{Omu}. These $\varpi_{L,R}$
coincide with those defined for supergravity in \reef{def:Sigmas} if
we take the decoupling limit in which $V_{L,R}\to\pm 1$.

\subsubsection{Superradiant bound}

We can easily derive from these formulas the bound on decay
frequencies for the CFT state dual to the black hole
\reef{cft:ExtErgo}. For this state, the left sector is at
temperature $T_L$ so the left-chirality operator $\O_L(x^+)$ gives
in \reef{Gtx} a thermal two-point function periodic in imaginary
time,
\beq \langle \O_L^\dagger(x^+)\O_L(0)\rangle_{T_L} \sim
\left(\frac{\pi T_L}{\sinh(\pi T_L x^+)}\right)^{2+\ell}\,.
\eeq
The
right sector is at zero-temperature, and so the boson gives the two
point function $\partial_-X(x^-) \partial_-X(0)\sim 1/(x^-)^2$ and
each fermion gives $ \psi(x^-) \psi(0) \sim 1/x^-$, so
\beq
\langle\O_R^\dagger(x^-)\O_R(0)\rangle_{0}\sim
\left(\frac{1}{x^-}\right)^{2+\ell}\,,
\eeq
and the integration over
the right sector in the decay rate \reef{decayG} gives a factor
\beq
\int dx^- e^{-i\varpi_R
x^-}\left(\frac{1}{x^--i\varepsilon}\right)^{2+\ell}\,.
\eeq
This
contour integral vanishes for $\varpi_R>0$, so
\beq
\Gamma_{\ell,p,m_{R,L}}(\omega) \propto \Theta(-\varpi_R)\,.
\eeq
This bound on frequencies coincides with the one we obtained from the
supergravity side, \reef{ExtLim:limSigma}, \reef{ExtLim:limrate}, in the
extremal limit where $\mu_R\to \Omega_R/2$ \reef{Omuext}, and in the
decoupling limit in which $V_H\to 1$. We feel, nevertheless, that the
microscopic derivation we gave in Sec.~\ref{sec:bound} is physically
more transparent.

\subsubsection{Absorption cross section: general case}

It is actually possible to compute the absorption cross section for
the more general case where both sectors are at temperatures $T_L$
and $T_R$ so we can compare it with the general results we obtained
from the supergravity side. We follow
\cite{Maldacena:1997ih,Gubser:1998ex,David:2002wn} but discuss the
general case with non-vanishing $\mu_{L,R}$ and $p$. The Green's
function \reef{Gtx} now has thermal correlation functions from both
sectors,
\begin{eqnarray}
  \label{Green}
   {\cal G}(t,x) = (-1)^\ell C_\O
    \left( {\pi T_L \over \sinh \left(\pi T_L x^+\right)} \right)^{2 +\ell}
    \left( {\pi T_R \over \sinh \left(\pi T_R x^-\right)} \right)^{2 +\ell},
\end{eqnarray}
where we include a factor $C_{\O} = C_{\O_+} C_{\O_-}$ that accounts
for the normalization factors of the operators, and which we will
discuss below jointly with the vertex factor $\mathcal{V}$.
The CFT absorption cross-section is the difference between
absorption and emission rates divided by the flux $\mathcal{F}$.
Then,
\begin{eqnarray}
  \label{SigmaCFT}
 \sigma^{CFT}_{\ell,p,m_{R,L}} &=& {\mathcal{V} \over {\cal F}}\,
    \int dx^+ dx^- e^{-i(\varpi_R x^-+\varpi_L x^+)} \left[ {\cal G}(t-i\epsilon,x) -
      {\cal G}(t+i\epsilon,x) \right]  \\
    &=& {C_\O \mathcal{V} \over {\cal F}} \,
       {(2 \pi T_L)^{1+\ell} (2 \pi T_R)^{1+\ell} \over
        \Gamma(2 +\ell)^2 } \nonumber\\
    && \hspace{1cm} \times \sinh\left(\frac{\varpi}{2T_H}\right)
      \left| \Gamma \left( \frac{\ell+2}{2} + i {\varpi_L \over  2\pi T_L} \right)
      \Gamma \left( \frac{\ell+2}{2} + i {\varpi_R \over  2\pi T_R} \right) \right|^2 \,. \nonumber
  \end{eqnarray}
Here $\varpi$ is exactly the same quantity that we introduced in the
supergravity analysis in \reef{def:Sigmas}, with $V_H$ given in
terms of $T_{L,R}$ as in \reef{VH}.

In order to find the factor $C_\O \mathcal{V}$, we first determine it for
s-wave ($\ell=0$) absorption by the 6D string (see \cite{David:2002wn}
for more details).
The minimally
coupled scalar $\Phi$ comes from an internally polarized graviton, say
$h_{67}$, so for $\ell=0$ the operator $\O$ is $\partial_- X^6_A \partial_+ X^7_A$, where the
index $A=1,\dots
N_1 N_5=Q_1 Q_5/R$ refers to the twist sector of the orbifold CFT.
Canonical normalization of the scalar field yields a factor $16\pi
G_6=8\pi^3 R$ in our units where $G_5=\pi/4$, times a factor of $2^2$
for the doubling due to $h_{67}=h_{76}$. On the other hand, we find a
conventional factor $1/16\pi^2$ from the two-point function of
$\partial_- X\partial_+ X$, and since we are
in the maximally twisted sector we must sum over all values of $A$. This determines
\beqa
(C_\O \mathcal{V})_{\ell=0}&=&4\times 8\pi^3 R\; \frac{Q_1 Q_5}{
16\pi^2 R}\nonumber\\
&=& 2\pi Q_1 Q_5\,.
\eeqa
When
$\ell>0$ the precise form of the vertex factor requires an explicit
derivation of the interaction vertex from string theory. We shall not
pursue this, but instead follow \cite{Mathur:1997et} to find
heuristically its dependence on all black hole parameters and all wave
quantum numbers except for $\ell$.
The $Q_1Q_5$ flavors
of open string fermions in the long string, combined along the boundary of the disk
diagram, yield a factor $(Q_1Q_5)^{\ell}$ for
the $\ell$ fermion pairs entering the interaction. We must also
divide it by $(\ell!)^2$ to account for the fact that we are
overcounting possibilities since the $\ell$ fermions in each sector
are indistinguishable. Additionally, the vertex must at least contain the
$\ell$ factors of momentum from the derivatives in it. Each
of the left and right fermions contribute, respectively, with
$(\omega\mp p)/2$ to this factor, yielding a total
\beq
C_\O \mathcal{V}=2\pi Q_1 Q_5
\left[\frac{1}{4}(\omega^2-p^2)\right]^{\ell}
\frac{(Q_1Q_5)^{\ell}}{(\ell!)^2} \hat{\mathcal{V}}_\ell\,, \eeq
where there remains an undetermined
$\ell$-dependent factor $\hat{\mathcal{V}}_\ell$, such that
$\hat{\mathcal{V}}_{\ell=0}=1$.

The flux
$\mathcal{F}$ measures the frequency or energy flow per unit cross
section. For a scalar of frequency $\omega$ and vanishing momentum
$p=0$ the canonically normalized flux of the incident field is
$\mathcal{F}=\omega$. However, if it has momentum $p$, then in the
frame of the string the frequency is increased by a Lorentz factor
$(1-p^2/\omega^2)^{-1/2}$, while the cross section is
Lorentz-contracted by $(1-p^2/\omega^2)^{1/2}$. Therefore, in
(\ref{SigmaCFT}) the flux is
\beq
\mathcal{F}=\frac{\omega}{1-p^2/\omega^2}\,.
\eeq
The final result
is then
\begin{eqnarray}
  \label{SigmaCFTf}
\sigma^{CFT}_{\ell,p,m_{R,L}} & = & \frac{8\pi\;
\hat{\mathcal{V}}_\ell}{(\ell!(\ell+1)!)^2}\frac{1}{\omega^3}\left(
\frac{\omega^2-p^2}{4}\right)^{\ell+1} \left(4 \pi^2 T_L T_R
Q_1Q_5\right)^{\ell+1} \sinh\left(\frac{\varpi}{2T_H}\right)\nonumber\\
&&     \times \left| \Gamma \left( \frac{\ell+2}{2} + i {\varpi_L
\over 2 \pi T_L} \right)
      \Gamma \left( \frac{\ell+2}{2} + i {\varpi_R \over 2 \pi T_R} \right)
      \right|^2\!.
  \end{eqnarray}

In order to compare this with the result \reef{absorptionSUGRA} from
supergravity, we must restrict the latter to the decoupling limit.
In this regime
\begin{eqnarray}
  \label{gasRelation}
  &&\mathcal{A}_H^{(5)} \rightarrow  4\pi^3 \frac{T_L T_R}{T_H} Q_1Q_5\,,
  \end{eqnarray}
while all the velocities become light-like, \reef{decV}, so
$\varpi_{L,R}$ and $\varpi$ are identical quantities in both sides
of the correspondence. Then we find
\begin{eqnarray}
 \sigma_{\ell,p,m_{R,L}}^{CFT}= \frac{\hat{\mathcal{V}}_\ell}{(\ell+1)^{2}}
 \;\sigma_{\ell,p,m_{R,L}}^{\mathrm{sugra}}\,.
 \label{CFTsugra}
\end{eqnarray}
So the decay rates agree remarkably well, and it would only remain to
check that a computation from first principles of
$\hat{\mathcal{V}}_\ell$, which is beyond the scope of this paper,
yields a perfect match. Taking the limit $T_R\to 0$ we find the decay
via superradiant emission of the ergo-cold black hole
\reef{ExtLim:limrate}.

\setcounter{equation}{0}\section{Conclusions}
\label{sec:Conclusions}

\bigskip

The recent progress in the microphysics of black holes is making it
clear that the gravitational description of a microscopic system
with a coarse-grain statistical degeneracy must exhibit a
horizon---this may require higher-derivative corrections if the
degeneracy scales too slowly with the mass. In this paper our aim
has been to find the microscopic origin of the ergoregion in
rotating black holes. Cold ergoregions provide a particularly clean
testbed, since they can only emit superradiant modes. What we have
found is that, in order for a superradiant ergoregion to be present,
the microscopic state must allow the annihilation of spin-carriers
to emit a bulk mode. If the system is at zero temperature, then
these spin-carriers must {\it necessarily} enter any interaction
leading to bulk emission. The superradiant frequency bound follows
then from the bound that the chemical potential sets on the energies
of excitation charged under its canonically-conjugated spin. All
these features are transparent in the $1+1$ CFTs we have considered,
in which the angular momentum is carried by fermions in (at least)
one sector, while the other sector {\it must} also contain some
excitations. This specific system has provided us with a simple,
elementary derivation of the superradiant bound \reef{sradbound}
without a detailed evaluation of absorption rates, which makes
manifest the fundamental role played by the Fermi-Dirac statistics
of the spin-carrying degrees of freedom.

It seems likely that the basic features of our microscopic picture
are also valid for any other gravitating object with a cold
ergoregion. The most familiar of these is the extremal Kerr black
hole. Ref.~\cite{Maldacena:1997ih} exhibited in a striking way how
the absorption rates from a Kerr black hole contained hints of a CFT
description. That this CFT must contain fermions as the carriers of
angular momentum seems difficult to dispense with, if one wants to
account for superradiant emission. Indeed, microscopic models for
the extremal Kerr and five-dimensional Myers-Perry black holes have
been proposed \cite{Emparan:2006it,Emparan:2007en,Horowitz:2007xq}.
These black holes are mapped, through symmetries and dualities, to
four-dimensional black holes of the kind we have discussed in
Section~\ref{subsec:4dbhs}. So the presence of superradiant emission
in these neutral black holes is understood, at least qualitatively,
in the same terms we have discussed: a filled Fermi sea in one
sector of the dual CFT. The quantitative recovery of the
superradiant bound is nevertheless not expected, since these neutral
black holes suffer non-trivial renormalizations of their masses and
energy levels (though not of their entropies) as a function of the
coupling.

Systems with cold ergoregions which are not U-dual to these black holes
are perhaps of more interest to test the applicability of our ideas
about the microphysics of superradiance. An instance of this are the
extremal rotating black rings with a dipole, in particular those in
which the dipole charge corresponds to a fundamental string and the
extremal limit is singular. The microscopic description of this dipole
ring has been described recently in \cite{BPEI}, and argued to possess
the right properties to fit our picture for a superradiating system: a
zero-temperature sector with angular momentum carriers, which can
interact with excitations from another sector and emit a spinning closed
string into the bulk. Note, though, that in the system in \cite{BPEI}
the angular momentum is not carried by a Fermi sea but by a bosonic
coherent state.

All these ergo-cold black holes provide, in a sense, cleaner
laboratories for the study of quantum emission from a black hole than do
non-extremal black holes. Since one of their sectors is in a ground
state, they are in a purer, less mixed state than non-extremal systems.
But still, their other sector is in a mixed, thermal ensemble. Therefore
it would be very interesting to consider states of the CFT such that
both sectors are in pure states but nevertheless they can interact and
decay by bulk emission. One such example is provided by the
non-supersymmetric smooth supergravity solitons with D1-D5-P charges in
\cite{ross}. On the microscopic side, they correspond to non-chiral
spectral flows of the Neveu-Schwarz ground state to non-BPS states in
the Ramond sector. The states have both sectors containing only
spin-carrying fermions. So we see that an interaction between the two
sectors will result into the emission of a spinning bulk scalar.
Following the overall picture proposed in this paper, superradiance is
naturally expected. Indeed, these supergravity solitons have ergoregions
(but not horizons) that have been shown to exhibit a superradiant
instability \cite{Cardoso:2005gj}. A correspondence between the two
pictures of the decay of precisely this type has been worked out in
detail very recently in \cite{Chowdhury:2007jx}, and conforms to the
overall ideas we have proposed.

\section*{Acknowledgments}

\noindent
 It is a pleasure to acknowledge stimulating discussions with Pau Figueras.
 This work has been supported in part by DURSI 2005 SGR 00082, CICYT
FPA 2004-04582-C02-02, FCT PTDC/FIS/ 64175/2006, and the European
Community FP6 program MRTN-CT-2004-005104. OJCD acknowledges
financial support provided by the European Community through the
Intra-European Marie Curie contract MEIF-CT-2006-038924. AM was
partially supported by a FPU grant from MEC (Spain).



\appendix

\section*{Appendix}

\setcounter{equation}{0}

\section{\label{sec:BHeffect}The near-horizon signature of superradiance}

It is natural to expect that the near-horizon geometry of the black
hole, which encodes in a dual manner the CFT description, contains
information about the possibility or not of superradiance. In the
dual CFT, superradiance refers to an interaction between the CFT and
a bulk scalar. The latter appears when the near-horizon geometry is
not fully decoupled from asymptotic infinity and therefore
disappears in the strict decoupling limit. Nevertheless, it would
seem natural that the near-horizon geometry could still encode a
signature that anticipates the existence of superradiant phenomena
in the full geometry. An effect of this kind was identified in
\cite{Bardeen:1999px} for the extremal Kerr black hole, which is the
simplest example of an ergo-cold black hole. From the study of
scalar propagation in the near-horizon geometry, they could indeed
identify an effect that signals superradiance in the Kerr solution.
In this appendix we show how this same effect is present in our
ergo-cold black hole, but not in the BMPV solution.

\subsection{\label{sec:NHmetric}Near-horizon geometry}

Take the  black hole solutions of the D1-D5-P system described in
(\ref{3charge}). To obtain their near-horizon geometry we introduce
\begin{eqnarray}
r^2=r_+^2+\varepsilon \rho \,, \qquad \tau=\gamma
\frac{t}{\varepsilon}\,,
 \label{NH:lambda}
\end{eqnarray}
where $\gamma$ is a constant to be defined later, and we take the
$\varepsilon\rightarrow 0$ limit. To avoid divergencies of the type
$1/\varepsilon$ and $1/\varepsilon^2$, this coordinate
transformation must be accompanied by a shift in the circle and
angular  directions,
\begin{eqnarray}
y=\tilde{y}+ {V_H} \,\frac{t}{\varepsilon}\,,\qquad
\phi=\tilde{\phi}+ \Omega_{\phi}\, \frac{t}{\varepsilon}\,,\qquad
\psi=\tilde{\psi}+ \Omega_{\psi} \,\frac{t}{\varepsilon}\,,
 \label{NH:shift}
\end{eqnarray}
where $\Omega_{\phi},\Omega_{\psi}, V_H$ represent the horizon
angular velocities defined in (\ref{omegaLR}), (\ref{V_H}). With the
shift (\ref{NH:shift}), the Killing vector $\partial/\partial_t$
becomes tangent to the horizon, \ie the new coordinates co-rotate
with the horizon. Next, we just write the near-horizon limit of the
extreme black hole metrics (in the end of this appendix we comment
on the non-extreme cases), since the near-horizon dilaton and RR
fields are not important for our discussion.

\begin{itemize}
\item Near-horizon geometry of the BPS black hole.

In this case one has $\Omega_{\phi,\psi}=0$ and
$\gamma=\ell_3^2\sqrt{Q_p}/2$ and one gets (dropping the
$\,\tilde{}\,$ in the angular coordinates),
\begin{eqnarray} \label{NH:BPSmetric}
ds_{NH}^2&=&\frac{\ell_3^2}{4}\left(-\rho^2
d\tau^2+\frac{d\rho^2}{\rho^2}\right)+\ell_3^2\left(
d\theta^2+\sin^2\theta d\phi^2 +\cos^2\theta
d\psi^2\right)+\frac{Q_p}{\ell_3^2} \left(
dy+\frac{\ell_3^2 \rho}{2\sqrt{Q_p}}d\tau\right)^2\nonumber \\
&& +\frac{2J_{\phi}}{\ell_3^2}\,dy\left( \sin^2\theta d\phi
+\cos^2\theta d\psi\right)\,,
\end{eqnarray}
where $\ell_3^2=\sqrt{Q_1Q_5}$.

\item Near-horizon geometry of the ergo-cold black hole.

One has $\Omega_{\phi}=\Omega_{\psi}$. We restrict our attention to
the simplest case with $a_1=a_2$. This case contains all the
features that are crucial for our study and does not loose any
important information, while avoiding non-insightful factors.

One gets, with
$\gamma=-\left[2a^{3}(c_{1}c_{5}c_{p}+s_{1}s_{5}s_{p})\right]^{-1}$
(and dropping the $\,\tilde{}\,$ in the angular coordinates),
\begin{eqnarray} \label{NH:ExtErgometric}
ds^{2}_{NH} & = &
\frac{K_{0}}{4}\left(-\rho^{2}d\tau^{2}+\frac{d\rho^{2}}{\rho^{2}}\right)+K_{0}
d \theta^{2}+K(\sin^{2}{\theta}d\phi+\cos^{2}{\theta}d\psi+P\rho
d\tau)^{2}\nonumber\\
& & +K_{0}\sin^{2}{\theta}(d\phi+P\rho
d\tau)^{2}+K_{0}\cos^{2}{\theta}(d\psi+P\rho
d\tau)^{2}\nonumber\\
& & +K_{y}\left[dy+K_{ty}\rho\ d\tau+P_{\phi
y}(\sin^{2}{\theta}d\phi+\cos^{2}{\theta}d\psi)\right]^{2} \,,
\end{eqnarray}
where $K_{0},K,P,K_{y},K_{ty},K_{\phi y}$ are constants given in
terms of the black hole parameters $a,\,\delta_{1,5,p}$ by
\begin{eqnarray}
K_{0} & = & 2a^{2}\sqrt{\cosh{(2\delta_{1})}\cosh{(2\delta_{5})}}\,,
\qquad
 K= \frac{2a^{2}[1-2\sech{(2\delta_{p})}(s_{1}s_{5}c_{p}-c_{1}c_{5}s_{p})^{2}]}{\sqrt{\cosh{(2\delta_{1})}\cosh{(2\delta_{5})}}}\,,\nonumber \\
P & = &
-\frac{1+\cosh{(2\delta_{1})}\cosh{(2\delta_{5})}+\cosh{(2\delta_{1})}\cosh{(2\delta_{p})}
+\cosh{(2\delta_{5})}\cosh{(2\delta_{p})}}{8(s_{1}s_{5}s_{p}+c_{1}c_{5}c_{p})^{2}}\,, \nonumber \\
K_{y}  &=& \frac{\cosh{(2\delta_{p})}}{\sqrt{\cosh{(2\delta_{1})}\cosh{(2\delta_{5})}}}\,, \nonumber \\
P_{\phi y} & = & -2a
(s_{1}s_{5}c_{p}-c_{1}c_{5}s_{p})\sech{(2\delta_{p})}\,, \qquad
P_{ty} = -a(s_{1}s_{5}c_{p}+c_{1}c_{5}s_{p}) \sech{(2\delta_{p})}\,.
\label{eq:NHconstants}
\end{eqnarray}
When $a_1\neq a_2$, there are overall $\theta$-dependent
multiplicative factors both on the $AdS_2$ and fibred $S^3$ parts of
the metric. They play no fundamental role in the analysis that we do
next.
\end{itemize}

The key observation in \reef{NH:ExtErgometric} is that the cross
terms between the time coordinate $\tau$ and the angular coordinates
$\phi,\psi$, are linear in the radial coordinate $\rho$ in the case
of the black hole with ergoregion. On the other hand, when the
ergoregion is absent, the radial dependence in the cross terms is
also not present. This feature plays an important role in the
near-horizon superradiant analysis that we do next.

\subsection{\label{sec:BHeffect3Q}The Bardeen-Horowitz signature of superradiance}

In this section we identify and justify the superradiant signature
in a near-horizon geometry. We refer to  this as the
Bardeen-Horowitz signature, since the feature that we will describe
was first identified by these authors in the extremal Kerr solution
\cite{Bardeen:1999px}. We will initially focus our analysis on the
near-horizon geometry (\ref{NH:ExtErgometric}) of the ergo-cold
black hole. We will single out the factor responsible for
superradiance in this case. Then we will observe that this factor is
absent when the ergoregion is not present, and in particular in the
BPS case.

Take (\ref{NH:ExtErgometric}). The following analysis gets
simplified if we carry dimensional reduction along $y$ (again we
will take waves with no momentum along the $T^4$, so this plays no
role in the discussion). This yields \footnote{We absorb a factor of
$K_{y}^{-1}$ in the lhs that comes from the KK dilaton (which being
constant plays no role): $ds^{2}_{NH(5)}\equiv
K_{y}^{-1}ds^{2}_{NH(5)}$. There is also a gauge field which is
irrelevant for our purposes, and whose components are
$A_{\tau}=K_{ty}\rho$, $A_{\phi}=K_{\phi y}\sin^{2}{\theta}$,
$A_{\psi}=K_{\phi y}\cos^{2}{\theta}$.}
\begin{eqnarray}\label{eq:5D_ansatz}
ds^{2}_{NH(5)} & = &
\frac{K_{0}}{4}\left(-\rho^{2}d\tau^{2}+\frac{d\rho^{2}}{\rho^{2}}\right)+K_{0}
d\theta^{2}+K(\sin^{2}{\theta}d\phi+\cos^{2}{\theta}d\psi+P\rho\ d\tau)^{2}\nonumber \\
& & +K_{0}\sin^{2}{\theta}(d\phi+P\rho\
d\tau)^{2}+K_{0}\cos^{2}{\theta}(d\psi+P\rho\ d\tau)^{2}\,.
\end{eqnarray}
This five-dimensional metric is of the form $AdS_{2}\times S^{3}$.
We can introduce global $AdS_{2}$ coordinates to cover the entire
spacetime \cite{Bardeen:1999px},
\begin{equation}\label{eq:global}
\rho=\sqrt{1+x^2}\,\cos{T}+x \,,\qquad \tau=\frac{\sqrt{1+x^2}
\,\sin T} {\rho}\,,
\end{equation}
whose ranges are $-\infty<T<\infty$, $-\infty<x<\infty$. To avoid
new crossed terms between $S^{3}$ and $AdS_{2}$ coordinates, we have
to shift $\phi$ and $\psi$ \cite{Bardeen:1999px},
\begin{equation}\label{eq:shift}
\phi,\psi=\tilde{\phi},\tilde{\psi}+P\log{\left[\frac{\cos{T}+x\sin{T}}{1+\sqrt{1+x^2}\,\sin{T}}\right]}\,.
\end{equation}
In these global coordinates the metric (\ref{eq:5D_ansatz}) reads,
\begin{eqnarray}\label{eq:5D_global}
ds^{2}_{NH(5)} & = &
\frac{K_{0}}{4}\left(-(1+x^{2})dT^{2}+\frac{dx^{2}}{1+x^{2}}\right)+K_{0}
d\theta^{2}+K(\sin^{2}{\theta}d\tilde{\phi}+\cos^{2}{\theta}d\tilde{\psi}+Px\, dT)^{2} \nonumber\\
& & +K_{0}\sin^{2}{\theta}(d\tilde{\phi}+Px\,dT)^{2}
+K_{0}\cos^{2}{\theta}(d\tilde{\psi}+Px\, dT)^{2}\,.
\end{eqnarray}

We now study the Klein-Gordon equation in this near-horizon
background (\ref{eq:5D_global}). Introducing the \emph{ansatz}
\begin{equation}\label{eq:scalar_ansatz}
\Phi=e^{-i(w T-m\tilde{\phi}-n\tilde{\psi})}\Theta(\theta)X(x)\,,
\end{equation}
the wave equation separates and yields
\begin{eqnarray}
&& \frac{1}{\sin{2\theta}}\frac{d}{d\theta}\left[\sin{2\theta}
\frac{d\Theta}{d\theta}\right]
+\left[\Lambda-\frac{m^{2}}{\sin^{2}{\theta}}
-\frac{n^{2}}{\cos^{2}{\theta}}\right]\Theta=0\,, \nonumber \\
&& \frac{d}{dx}\left[(1+x^{2})\frac{dX}{dx}\right]+\frac{1}{4}
\left[\frac{4[w+(m+n)Px]^{2}}{1+x^{2}}+\frac{K}{K+K_{0}}(m+n)^{2}
-\Lambda\right]X=0\,, \label{NHeq:radial}
\end{eqnarray}
where $K_{0}$, $K$ and $P$ are defined in
(\ref{eq:NHconstants}).\footnote{The separation constant is exactly
$\Lambda=\ell (\ell+2)$ (this is a consequence of working with the
$a_{1}=a_{2}$ case), and poses a bound on the other angular quantum
numbers: $\ell\ge|m|+|n|$.}

The radial equation presents an important feature. Indeed, apart
from the contribution coming from the piece $(m+n)Px$, this radial
equation is very similar to the equation describing perturbations in
a pure $AdS_2$ background \cite{Bardeen:1999px}. That is, in
(\ref{NHeq:radial}) we have $[w+(m+n)Px]^{2}$ instead of $w^{2}$
that is present in the pure $AdS_2$ case. The origin of this factor
can be easily traced back and found to be due to the presence of the
terms $P\rho d\tau$ in (\ref{NH:ExtErgometric}); see discussion at
the end of Section \ref{sec:NHmetric}. We next discuss the
implications of this property for the near-horizon signature of
superradiance.

In a WKB approximation the effective wavenumber for traveling waves
obeying (\ref{NHeq:radial}), $k=-\frac{i}{X}\frac{dX}{dx}$, is
\begin{equation}\label{eq:k}
k=\pm\frac{1}{4\sqrt{1+x^2}}\left[\frac{4[w+(m+n)Px]^{2}}{1+x^{2}}
+\frac{K}{K+K_{0}}(m+n)^{2}-\Lambda\right]^{1/2}\,
\end{equation}
from which follows the associated group velocity,
\begin{equation}\label{eq:group}
\frac{dw}{dk}=\pm\frac{4(1+x^{2})^{3/2}}{w+(m+n)Px}
\left[\frac{[w+(m+n)Px]^{2}}{1+x^{2}}+\frac{K}{K+K_{0}}(m+n)^{2}-\Lambda\right]^{1/2}\,.
\end{equation}
On the other hand, the phase velocity of the waves is $w/k$. As
first observed in \cite{Bardeen:1999px}, in the context of the Kerr
geometry, here the group and phase velocities can have opposite
signs. For positive $(m+n)P$ this occurs when $x<\frac{w}{(m+n)P}$,
while for negative $(m+n)P$ this is true when $x>\frac{w}{(m+n)P}$.
An original argument from Press and Teukolsky
\cite{PressTeukolsky:II}, concludes that this defines the
near-horizon superradiant regime. Indeed, the opposite sign between
group and phase velocities of a wave in the vicinity of a horizon is
responsible for the fundamental origin of superradiance.
Classically, only ingoing waves are allowed to cross the horizon.
The quantity that defines the physical direction of a wave is its
group velocity rather than its phase velocity. So the classical
absorption of incident waves is described by imposing a negative
group velocity as a boundary condition. Note however that in the
near-horizon superradiant regime above mentioned, the associated
phase velocity is positive and so waves appear as outgoing to an
inertial observer at spatial infinity. Thus, energy is in fact being
extracted, \ie superradiance is active \cite{PressTeukolsky:II}.

At this point, we make a contact with the other extreme case and
with the discussion in the end of subsection \ref{sec:NHmetric}. For
the BPS black hole, there is no radial dependence in the cross terms
between the time and angular coordinates in its near-horizon
geometry (\ref{NH:BPSmetric}). As a consequence, there is no linear
term in the frequency in the wave equation associated with this
background. But this implies that group and phase velocities always
have the same sign in this background. Thus there is no available
room for a superradiant regime in the near-horizon geometries of
extreme black holes without ergoregion. Finally note that in a
general non-extreme black hole the situation is quite similar to the
ergo-cold black hole  in what concerns the issue discussed in this
appendix.


\end{document}